
\voffset=-1.5truecm\hsize=16.5truecm    \vsize=24.truecm
\baselineskip=14pt plus0.1pt minus0.1pt \parindent=12pt
\lineskip=4pt\lineskiplimit=0.1pt      \parskip=0.1pt plus1pt

\def\st{\scriptstyle}


\let\a=\alpha \let\b=\beta  \let\d=\delta \let\e=\varepsilon
 \let\g=\gamma     \let\k=\kappa 
\let\m=\mu \let\n=\nu      
\let\r=\rho \let\s=\sigma  \let\th=\vartheta
  \let\z=\zeta
\let\D=\Delta  \let\G=\Gamma \let\L=\Lambda


\global\newcount\numsec\global\newcount\numfor
\gdef\profonditastruttura{\dp\strutbox}
\def\senondefinito#1{\expandafter\ifx\csname#1\endcsname\relax}
\def\SIA #1,#2,#3 {\senondefinito{#1#2}
\expandafter\xdef\csname #1#2\endcsname{#3} \else
\write16{???? il simbolo #2 e' gia' stato definito !!!!} \fi}
\def\etichetta(#1){(\veroparagrafo.\veraformula)
\SIA e,#1,(\veroparagrafo.\veraformula)
 \global\advance\numfor by 1
 \write16{ EQ \equ(#1) ha simbolo #1 }}
\def\etichettaa(#1){(A\veroparagrafo.\veraformula)
 \SIA e,#1,(A\veroparagrafo.\veraformula)
 \global\advance\numfor by 1\write16{ EQ \equ(#1) ha simbolo #1 }}
\def\BOZZA{\def\alato(##1){
 {\vtop to \profonditastruttura{\baselineskip
 \profonditastruttura\vss
 \rlap{\kern-\hsize\kern-1.2truecm{$\scriptstyle##1$}}}}}}
\def\alato(#1){}
\def\veroparagrafo{\number\numsec}\def\veraformula{\number\numfor}
\def\Eq(#1){\eqno{\etichetta(#1)\alato(#1)}}
\def\eq(#1){\etichetta(#1)\alato(#1)}
\def\Eqa(#1){\eqno{\etichettaa(#1)\alato(#1)}}
\def\eqa(#1){\etichettaa(#1)\alato(#1)}
\def\equ(#1){\senondefinito{e#1}$\clubsuit$#1\else\csname e#1\endcsname\fi}


\def\bb{\hbox{\vrule height0.4pt width0.4pt depth0.pt}}\newdimen\u
\def\pp #1 #2 {\rlap{\kern#1\u\raise#2\u\bb}}

\def\ins #1 #2 #3 {\rlap{\kern#1\u\raise#2\u\hbox{$#3$}}}

\def\pallina{{\kern-0.4mm\raise-0.02cm\hbox{$\scriptscriptstyle\bullet$}}}
\def\palla{{\kern-0.6mm\raise-0.04cm\hbox{$\scriptstyle\bullet$}}}
\def\pallona{{\kern-0.7mm\raise-0.06cm\hbox{$\displaystyle\bullet$}}}


\def\data{\number\day/\ifcase\month\or gennaio \or febbraio \or marzo \or
aprile \or maggio \or giugno \or luglio \or agosto \or settembre
\or ottobre \or novembre \or dicembre \fi/\number\year}

\setbox200\hbox{$\scriptscriptstyle \data $}

\newcount\pgn \pgn=1
\def\foglio{\number\numsec:\number\pgn
\global\advance\pgn by 1}
\def\foglioa{a\number\numsec:\number\pgn
\global\advance\pgn by 1}

\footline={\rlap{\hbox{\copy200}\ $\st[\number\pageno]$}\hss\tenrm \foglio\hss}


\def\sqr#1#2{{\vcenter{\vbox{\hrule height.#2pt
\hbox{\vrule width.#2pt height#1pt \kern#1pt
\vrule width.#2pt}\hrule height.#2pt}}}}
\def\square{\mathchoice\sqr34\sqr34\sqr{2.1}3\sqr{1.5}3}

\let\ciao=\bye 
\def\pagina{{\vfill\eject}} \def\\{\noindent}

  \let\io=\infty 

\let\dpr=\partial

\def\RR{{\cal R}}\def\SS{{\cal S}} 
 \def\AA{{\cal A}}

\def\tende#1{\vtop{\ialign{##\crcr\rightarrowfill\crcr
              \noalign{\kern-1pt\nointerlineskip}
              \hskip3.pt${\scriptstyle #1}$\hskip3.pt\crcr}}}
\def\otto{{\kern-1.truept\leftarrow\kern-5.truept\to\kern-1.truept}}
\def\arm{{}}
\font\smfnt=cmr8 scaled\magstep0


\vglue.5truecm
{\centerline{\bf CRITICAL $({\bf\Phi}^{4})_{3,\>\epsilon}$}}
\vglue1.truecm
{\centerline{{\bf D. C. Brydges}$^{1,}$\footnote{${}^*$}
{\arm Partially supported by NSERC of Canada}, 
{\bf P. K. Mitter}$^{2,}$\footnote{${}^{**}$}
{\arm Laboratoire Associ\'e au CNRS, UMR 5825}, 
{\bf B. Scoppola}$^{3,}$\footnote{${}^{***}$}
{\arm Partially supported by CNR, G.N.F.M. and MURST}}}
\vglue.5truecm
{\centerline{\smfnt 
$^1$Department of Mathematics, University of British Columbia}}
{\centerline{\smfnt 121-1984 Mathematics Rd, Vancouver, B.C, Canada V6T 1Z2}}
{\centerline{\smfnt e-mail: db5d@math.ubc.ca}}
\vglue.5truecm
{\centerline{\smfnt $^2$Laboratoire de Physique math\'ematique, 
Universit\'e Montpellier 2}}
{\centerline{\smfnt Place E. Bataillon, Case 070, 
34095 Montpellier Cedex 05 France}}
{\centerline{\smfnt e-mail: mitter@LPM.univ-montp2.fr}}
\vglue.5truecm
{\centerline{\smfnt $^3$Dipartimento di Matematica, 
Universit\'a ``La Sapienza'' di Roma}}
{\centerline{\smfnt Piazzale A. Moro 2, 00185 Roma, Italy}}
{\centerline{\smfnt e-mail: benedetto.scoppola@roma1.infn.it}}
\vglue2.truecm

{\centerline{ABSTRACT}}
\vglue.5truecm
\\The Euclidean $(\phi^{4})_{3,\>\epsilon}$ model in ${\bf R}^3$ corresponds 
to a perturbation by a $\phi^4$ interaction
of a Gaussian measure on scalar fields with a covariance
depending on a real parameter $\e$ in the range $0\le \e \le 1$. For
$\e =1$ one recovers the covariance of a massless scalar field in
${\bf R}^3$. For $\e =0$ $\phi^{4}$ is a marginal interaction. 
For $0\le \e < 1$ the covariance continues to be Osterwalder-Schrader and
pointwise positive. After introducing cutoffs we prove that for $\e > 0$, 
sufficiently small, there exists a non-gaussian fixed point ( with one unstable
direction) of the Renormalization Group iterations. These iterations converge 
to the fixed point on its stable (critical) manifold which is constructed.

\pagina

\numsec=1\numfor=1

\vskip1truecm
\noindent{\bf 1. INTRODUCTION, MODEL, RG TRANSFORMATION } 
\vskip0.5truecm
\noindent{\it 1.1 Introduction}
\vskip0.5truecm
\\Let $\phi$ be a mean zero Gaussian scalar random field on ${\bf R}^3$
with covariance 
$$
E(\phi(x)\phi(y))={{\rm const}\over |x-y|^{(3-\epsilon)/2 }}$$ 
$$ = \int {d^{3}p\over (2\pi)^{3}} e^{ip\cdot(x-y)}
(p^{2})^{-(3+\epsilon)/4} \Eq(1.1)
$$ 
Here $p^2=\vert p\vert^2$ is the standard Euclidean norm in ${\bf R}^3$ and
$p\cdot x$ is the standard scalar product.
$\e$ is a real parameter which we take in the region
$0\le \epsilon <1$. Note that for $\epsilon =1$ we have the standard 
massless free scalar field in ${\bf R}^3$.
\vskip0.3truecm
\\This covariance  $(-\Delta)^{-(3+\epsilon)/4}$ has interesting physical properties. For example, Osterwalder-
Schrader positivity plays no role in the present paper, but it is of
interest because scaling limits of these theories will be Euclidean
quantum field theories. $(-\Delta)^{-(3+\epsilon)/4}$ is  Osterwalder-
Schrader positive as well as being pointwise positive not only for $\epsilon=1$
but also in the range that
we consider, namely $0\le \epsilon <1$. 
In the latter range we have the 
convergent integral representation

$$(-\Delta)^{-(3+\epsilon)/4}(x-y)=1/c_{\epsilon} \int_0^{\infty}ds\ \ 
s^{-(3+\epsilon)/4}(-\Delta + s)^{-1}(x-y) \Eq(1.2)  $$
where

$$c_{\epsilon}=\int_0^{\infty}ds\ \ s^{-(3+\epsilon)/4}(1+s)^{-1} \Eq(1.3)$$
The Osterwalder-Schrader and pointwise positivities now follow from those
of $(-\Delta + s)^{-1}(x-y)$. 
\vskip0.3truecm
\\Furthermore $(-\Delta)^{-(3+\epsilon)/4}$ is the Green's function (or potential) 
for a stable L\'evy process in ${\bf R}^3$  with parameters $(\alpha, \beta)$
in the L\'evy-Khintchine notation [KG], with 
the characteristic exponent $\alpha =(3+\epsilon)/2$, and $\beta =0$.
This process has jumps and diffuses very fast.  This property also
plays no role in the present paper but we expect that self-avoiding
Levy processes are accessible to the methods of this paper.

\vskip0.3truecm
\\These properties make the study of the above gaussian random field and its
non-linear perturbations worthwhile. In particular we will study here the
critical properties of a model corresponding to the  partition function 
$${\bf Z} =\int d\mu_C(\phi)\ \  e^{-V_{0}(\phi)} \Eq(1.4)$$
where $d\mu_C$ is the Gaussian measure with covariance $C$ and $C$ is
$(-\Delta)^{-(3+\epsilon)/4}$ with an ultraviolet cutoff 
described below and 
$$V_{0}(\phi)=V(\phi,C,g_{0},\mu_{0})=g_{0}\int d^3x:\phi^4:_C(x)+
\mu_{0}\int d^3x:\phi^2:_C(x) \Eq(1.5)$$
and the coupling constant $g$ is held strictly positive.
Moreover in order to define the 
model completely we must also introduce a volume cutoff. These cutoffs will
be introduced presently when we give a precise definition of the covariance 
$C$ in\equ(1.5) as well as that of the model. 

\\From\equ(1.1) we can read off the canonical scaling dimension 
$[\phi ]$ of $\phi$
$$[\phi] =(3-\epsilon)/4 \Eq(1.5.1)$$
This, together with\equ(1.4), implies that we can assign to $g,\mu$  the 
dimensions 
$$[g]=\epsilon,\ \  [\mu]= (3+\epsilon)/2 \Eq(1.6)$$
Note that we have not put in a term
$${z\over 2}\int d^3x|\nabla\phi(x)|^2$$
in \equ(1.5). This is because the dimension $[z]= -(1-\e)/2$. 
Hence for $\e<1$ this is a candidate for an irrelevant (stable) direction. 

\\ We thus expect from Wilson's theory of critical phenomena [KW] that for
$\epsilon >0$ the critical (infra-red) properties of the model are dominated
by a non-Gaussian fixed point of Renormalization Group iterations with 
$g=g_*=O(\epsilon)$ provided the unstable parameter $\mu$ 
is fine tuned to a critical function $\mu_c(g)$ which determines the 
stable (critical) manifold of the fixed point.
In the present work we will prove the existence of the non-Gaussian fixed 
point for $\epsilon >0$  held sufficiently small and, on the way, construct
the stable manifold. 
\vskip0.3truecm

The mathematical analysis of Renormalization Group (henceforth denoted
RG) transformations has by now a long history [F,BG]. Our particular
line of attack is influenced by a series of works which started with
[BY], developed further in [DH1,DH2, BDH-est,BDH-eps], with more
recent developments in [MS]. We shall be concerned here with these
latter developments. In [MS] fluctuation covariances of finite range
were exploited, and this simplifies considerably the RG analysis. In
particular the analysis of the fluctuation integration becomes a
matter of geometry and one no longer needs the cluster expansion and
analyticity norms. In the continuum approach of [MS] that is also
adopted here the existence of fluctuation covariances with finite
range follows easily from a judicious choice of a class of ultraviolet
cutoffs. This raises the general question of which gaussian random
fields can be decomposed into sums of fluctuation fields with
covariances with finite range.  A partial answer which includes the
standard massless Euclidean field with lattice cutoff will be given in
[Gu].  Only the existence of multiscale decompositions with the above
finite range property (together with some regularity and positivity
properties) is required in what follows.  \vskip0.3truecm The present
work borrows many technical considerations introduced in
[BDH-est]. Although these are independent of the manner of
treating the fluctuation step, we repeat some of them because the
simpler norms in this paper allow easier proofs.  We also borrow some
ideas from [BDH-eps] where a related model (which however does not
possess the physical properties mentioned earlier) was considered and
the existence of a non-Gaussian fixed point was proved.  We use this
paper as an opportunity to improve previous arguments. In
particular, in Section 4, there are much simpler formulas for the
remainder after approximating the RG step by second order perturbation
theory.

\vskip0.3truecm
We plan to study in a subsequent 
work critical properties of Self-Avoiding L\'evy
Flights in ${\bf Z}^3$ for the L\'evy-Khintchine parameters given above with 
$\alpha > \alpha_c $ and $\alpha -\alpha_c$ very small. Here $\alpha_c = 3/2$
at which value we expect ( heuristically) mean field behaviour.

\vskip0.5truecm
\noindent{\it 1.2  Multiscale decomposition}
\vskip0.5truecm
\\We introduce a special type of ultraviolet cutoff as follows: 
Let $g$ be a non-negative, $C^{\infty}$, $O(3)$ invariant function
of compact support in ${\bf R^3}$ such that $g(x)$ vanishes for
$|x|\ge 1/2$. Define $u=g*g$. Thus $u$ is positive, $C^{\infty}$, and
of compact support: $u(x)=0$ for $|x|\ge 1$. First we note note that 
$$
{{\rm const}\over |x-y|^{(3-\epsilon)/2 }} =
\int_{0}^{\infty}{dl\over l}\ \ l^{-(3-\epsilon)/2}\ \ 
u({x-y\over l})
$$
by scaling the variable of integration. Define


$$C(x-y)=\int_{1}^{\infty}{dl\over l}\ \ l^{-(3-\epsilon)/2}\ \ 
u({x-y\over l}) \Eq(1.10)$$
Because the lower limit is $1$ and not $0$ this $C$ is pointwise positive and 
$C^{\infty}$. 
\vskip0.3truecm
\noindent{\it Remark}

\\We can exhibit C in the traditional form with an ultraviolet cutoff function. 
Let $\hat u$ be the Fourier transform of $u$. Because of $O(3)$ invariance we 
can write ${\hat u}(p)=v(p^2)$. Then it is easy to see from the above that

$$C(x-y)=\int {d^{3}p\over (2\pi)^{3}} e^{ip.(x-y)} F(p^{2}) 
(p^{2})^{-(3+\epsilon)/4} \Eq(1.11)$$
where

$$F(p^{2})=1/2\int_{p^2}^{\infty}{ds\over s}\ \ s^{(3+\epsilon)/4}\ \
v(s) \Eq(1.12)$$

\\From this we see that F is positive, continuous and of fast decrease 
( since $v(p^2)=|{\hat g}(p)|^2$ and $g$ is of compact support) and can be 
thus identified as the ultraviolet cutoff function.

\vskip0.5truecm
\\Let $L\ge 2$ be an integer. Let
$$\Gamma (x-y)=\int_{1}^{L}{dl\over l}\ \ l^{-(3-\epsilon)/2}\ \ 
u({x-y\over l}) \Eq(1.8)$$
Clearly $\Gamma$ is $C^{\infty}$, pointwise positive and of finite range: 
$$\Gamma(x-y)=0 :\ \ |x-y|\ge L \Eq(1.9)$$
$\Gamma$ is our {\it fluctuation covariance} and it satisfies
$$C(x-y)=\Gamma(x-y) + L^{-(3-\epsilon)/2} C({x-y\over L}). 
\Eq(1.17)$$

\\Moreover because of our choice of the form of $u$, namely $u=g*g$,
$\G$ and $C$ are generalized positive definite. Now it can be shown
(see e.g. Section 1.1 of [MS]) that under these conditions for
any compact set $\Lambda_N$ in $\bf R^3$    there
exists a Gaussian measure $d\mu_{C}$ of mean $0$ and covariance $C$
supported on the Sobolev space $H_{s}(\Lambda_N)$ for any
$s>0$. Choosing $s > 3/2 +2$ is sufficient for our purpose. Then, by
Sobolev embedding, sample paths are at least twice
differentiable.  \vskip0.5truecm

\vskip0.3truecm
\\Define the compact set
$$\Lambda_N =[-{1\over 2}L^{N},{1\over 2}L^{N}]^3\subset \bf R^3.
\Eq(1.13)$$
Our model with an ultraviolet cutoff and volume cutoff is now defined by
the partition function
$${\bf Z} =\int d\mu_{C}(\phi)\ \ {\cal Z}_{0}(\Lambda_{N},\phi) 
\Eq(1.15)$$
where
$${\cal Z}_{0}(\Lambda_{N},\phi)=e^{-V(\Lambda_{N},\phi)} \Eq(1.16)$$
and the potential has now been restricted to the volume $\Lambda_{N}$. 
The Wick ordering in the potential is with respect to the infinite
volume covariance $C$ and this is well defined because the Wick constants
are finite.

\vskip0.5cm
\noindent{\it 1.3 Renormalization Group transformation}
\vskip0.5truecm
\\From\equ(1.17), 
if we now define the rescaled field
$$\RR\phi(x) =\phi_{L^{-1}}(x) =L^{-(3-\epsilon)/4}\phi(x/L)\Eq(1.18)$$
we get 
$$\int d\mu_{C}(\phi){\cal Z}_0(\Lambda_N,\phi)=
\int d\mu_{C}(\phi){\cal Z}_1(\Lambda_{N-1},\phi)\Eq(1.18.1)$$
where
$${\cal Z}_1(\Lambda_{N-1},\phi)=
\int d\mu_{\Gamma}(\zeta){\cal Z}_0(\Lambda_N,\zeta +\phi_{L^{-1}} )
\Eq(1.19)$$
The iteration of\equ(1.19) will constitute our RG transformations. After $n$
steps we have 
$$\int d\mu_{C}(\phi){\cal Z}_0(\Lambda_N,\phi)=
\int d\mu_{C}(\phi){\cal Z}_n(\Lambda_{N-n},\phi)\Eq(1.20)$$
where
$${\cal Z}_n(\Lambda_{N-n},\phi)=
\int d\mu_{\Gamma}(\zeta){\cal Z}_{n-1}(\Lambda_{N-n+1},\zeta +\phi_{L^{-1}} )
\Eq(1.21)$$
After $N-1$ steps we arrive at ${\cal Z}_{N-1}(\Lambda_{1},\phi)$  where 
$\Lambda_{1}$ is the $L$- block $[-{1\over 2}L,{1\over 2}L]^3$.
\vskip0.3true cm
\\In order to analyse the RG transformations, it is convenient to write the
partition function density in a {\it polymer gas } representation. Pave
${\bf R}^3$ with closed unit blocks denoted henceforth $\Delta$.
Now let $\Lambda\subset {\bf R}^3$ 
be the volume after a certain number of RG steps. We take $\Lambda$ with
the induced paving. A {\it connected} polymer $X$ is 
a connected union of a subset of these closed unit blocks and is thus closed.
A polymer activity $K(X,\phi)$ is a map $X,\phi\rightarrow {\bf R}$ where
the fields $\phi$ depend only on the points of $X$. We shall only consider
polymer activities supported on connected polymers. This notion will
be preserved by the RG operations. We then write, suppressing the dependence
on $\phi$,
$${\cal Z}(\Lambda)=\sum_{N=0}^{\infty}{1\over N!}e^{-V(\Lambda\backslash X)}
\sum_{X_1,..,X_N} \prod_{j=1}^N K(X_j) \Eq(1.22)$$
where the connected polymers $X_j$ are disjoint, $X=\cup_1^N X_j$ and 
$V(Y)=V(Y,\phi,C,g,\mu)$ is given by\equ(1.5) with parameters $g,\mu$ and 
integration over $Y$. Initially of course K is absent, but they are naturally 
generated by the RG operations and the above form is stable.

\vglue.3truecm 
\\It is possible to rewrite the above in a more compact form if we
extend the polymer algebra by {\it cells} as done in [BY]. A {\it
cell} may be the interior of a block, an open face, an open edge or a
vertex. A polymer, which is a union of closed blocks, can be uniquely
written as a disjoint union of cells, but the point is that all other
sets generated by our manipulations, such as complements of polymers,
can also be uniquely written as disjoint unions of cells. We define a
commutative product, denoted $\circ$, on functions of sets (unions of
cells), in the following way $(F_1\circ
F_2)(X)=\sum_{Y,Z:Y{\circ\atop\cup}Z=X}F_1(Y)F_2(Z)$ where
$X=Y{\circ\atop\cup}Z$ iff $X=Y\cup Z$ and $Y\cap Z=\emptyset$.  The
$\circ$ identity ${\cal I}$ is defined by ${\cal I}(X)=1$ if
$X=\emptyset$ and otherwise vanishes. Finally we can define an ${\cal
E}$xponential operation on functions of unions of cells as the usual
power series based on the $\circ$ product and the definition of the
$\circ$ identity ${\cal I}$. This has the properties of the usual
exponential. We also define a {\it space filling} function $\square$
by ${\square}(X) =1$ if $X$ is a cell and otherwise vanishes.  We then
have ${\cal E}xp\> \square =1$ and it easy to see that\equ(1.22) can
be rewritten as
$${\cal Z}(\Lambda)=[{\cal E}xp (\square\> e^{-V}+K)](\Lambda) 
\Eq(1.23)$$

\vskip.5truecm
\\{\it The Formal Infinite Volume Limit}
\vskip.2truecm

\\By writing the integrands in the form \equ({1.23}), the $i$th RG
transformation induces a map 

$$f_{N-i}:(K_{i-1},V_{i-1}) \mapsto
(K_{i},V_{i})$$

\\which will be described in detail in the next
section. The subscript $N-i$ is there because this map has
dependence on the region $\Lambda_{N-i}$.  For any set $X$, 

$$V_{i}(X)= \sum_{\Delta \subset X}V_{i} (\Delta )$$

\\so $V_{i}$ is determined by $\{V_{i} (\Delta): \Delta \subset
\Lambda_{N-i}\}$.  Our formulas for the RG map will show that for any
fixed set $X\subset \Lambda_{N-i}$, the polymer activity $K_{i}
(X,\phi )$ is independent of $N$ for all $N$ large enough so that
$\Lambda_{N-i}$ contains $X$ and likewise $V_{i} (\Delta)$ is
independent of $N$ for all $N$ large enough so that $\Lambda_{N-i}$
contains $\Delta$ and a neighborhood of $\Delta$.  A detailed look at
our formulas, particularly the section on extraction, shows that the
neighborhood has diameter $8$.  {\it $\lim_{N\rightarrow \infty}K_{i}
(X)$ and $\lim_{N\rightarrow \infty}K_{i} (\Delta)$ exist and, in this
pointwise sense, the infinite volume limit $f=\lim_{N\rightarrow
\infty}f_{N-i}$ exists}.  In this paper we prove that $f$ has a fixed
point in a Banach space of polymer activities $K$.

\vglue.3truecm 
\\The finite volume RG could also be studied by these methods by
including in $V$ a surface integral over the boundary of
$\Lambda_{N-i}$ which fits naturally in this scheme as an object
associated to $D-1$ cells on the boundary.

\vskip0.5truecm
\numsec=2\numfor=1
\noindent{\bf 2. REGULATORS, DERIVATIVES AND NORMS }
\vskip0.5truecm
\noindent
{\it 2.1 Regulators} 
\vskip0.5truecm
\\We first introduce a {\it large field regulator} which measures the
growth of polymer activities in the fields $\phi$, actually in  $\dpr \phi$:
$$G_{\kappa}(X,\phi)=e^{\kappa\Vert\phi\Vert^2_{X,1,\sigma}} \Eq(2.1)$$
where
$$\Vert\phi\Vert^2_{X,1,\sigma}=\sum_{1\le |\alpha|\le\sigma}\Vert
\partial^\alpha\phi\Vert^2_X \Eq(2.2)$$
Here $\Vert\phi\Vert_X$ is the $L^2$ norm and $\alpha$ is a multi-index.
We take $\sigma > 3/2 +2$
so that this norm can be used in Sobolev inequalities to control
$\phi$ and its first two derivatives pointwise. After the function $u$
is fixed, the parameter $\kappa
> 0$ is fixed, for the whole paper, by a choice depending only on $u$,
so that for all $L\ge 1$
the large field regulator satisfies the {\it stability property}
$$\int d\mu_{\Gamma}(\zeta)\> G_{\kappa}(X,\zeta +\phi)\le 2^{|X|}G_{2\kappa}
(X,\phi) \Eq(2.3)$$
where $X$ is a polymer and $|X|$ is the number of unit blocks in $X$.
This can be shown in the same way as in the proof of the stability property
of the large field regulator in Section 2.4 of [BDH-est]. 

\\Now hold $L$ sufficiently large and recall that $\epsilon <1$.
Then we get after rescaling
$$\int d\mu_{\Gamma}(\zeta)\> G_{\kappa}(X,\zeta +\phi_{L^{-1}})\le 2^{|X|}
G_{\kappa}(L^{-1}X,\phi) \Eq(2.3.1)$$
This follows easily from the scaling property\equ(1.18) of the fields $\phi$ 
which gives 
$$\Vert\phi_{L^{-1}}\Vert^2_{X,1,\sigma}\le L^{-(1-\epsilon)/ 2}
\Vert\phi\Vert^2_{L^{-1}X,1,\sigma}$$
\vskip0.5truecm
\noindent
Next we introduce a {\it large set regulator}. Let $X$ be a connected
polymer. We define
$$\AA_{p} (X)= 2^{p|X|}L^{(D+2)|X|} \Eq(2.4)$$
where for us the dimension of space $D=3$, and $p\ge 0$ is an integer.
\vskip0.3truecm
\\Call a connected polymer {\it small} if $|X|\le 2^D$. A connected polymer
which is not small is called {\it large}.
\noindent
Let $\bar{X}^{L}$ be the $L$-closure of $X$. This is the smallest union of 
$L$-blocks containing $X$. Let $L$ be sufficiently large. 
Then we have from Lemma~1 of [BDH-est] the following two facts:

\noindent
For any connected polymer $X$ and for any integer $p\ge 0$
$$\AA (L^{-1}\bar{X}^{L}) \le c_{p}\AA_{-p}(X) \Eq(2.5)$$
For $X$ a {\it large} connected polymer,
$$\AA (L^{-1}\bar{X}^{L}) \le c_{p}L^{-D-1}\AA_{-p}(X) \Eq(2.6)$$ 
Here $c_p =O(1)$ is a constant independent of $L$.
\vskip0.5truecm
\noindent
{\it 2.2 Derivatives in fields and polymer activity norms}
\vskip0.5truecm
The following definitions are motivated by those in [BDH-est] the
main difference being that we will need only a finite number of field
derivatives and for them only {\it natural} norms. The kernel norms
defined below are different.
\vskip0.3truecm
\noindent
For a polymer activity $K(X,\phi)$ define the $n$-th derivative in the
direction $(f_{1},...,f_{n})$ as:
$$(D^{n}K)(X,\phi ;f^{\times n})=\prod_{j=1}^{n}(\partial/ \partial s_j)
\>K(X,\phi +\sum s_{j}f_{j})\mid_{ s_j  =0\>\forall j} \Eq(2.7)$$
where the shorthand notation $f^{\times n}$ stands for $f_1,f_2,...,f_n$.    
The functions $f_j$ belong to $C^{2}(X)$ and we assume that such derivatives 
exist for $n\le n_0$ for some $n_0$.

\noindent
We will measure the size of such derivatives, which are multilinear
functionals on $C^{2}(X)$, by the norm:
$$\Vert(D^{n}K)(X,\phi)\Vert=\sup_{\Vert f_j\Vert_{C^2 (X)}\le 1\>\forall j}\>
|(D^{n}K)(X,\phi ;f^{\times n})| \Eq(2.8)$$
\vskip0.3truecm
\\Let $h>0$ be a real parameter. We  define the following set of norms:
$$\Vert K(X,\phi)\Vert_h =\sum_{j=0}^{n_0} {h^{j}\over j!}\Vert(D^{j}K)(X,\phi)\Vert \Eq(2.9)$$
and then
$$\Vert (K(X)\Vert_{h,G_{\k}}=\sup_{\phi\in C^2 (X)}\Vert K(X,\phi)\Vert_h
\>G_{\kappa}^{-1}(X,\phi) \Eq(2.10)$$

\\These norms differ from norms used in [BDH-est] in having the
supremum over $\phi$ outside the sum over derivatives, as well as
involving only finitely many derivatives. They are easier to use and
retain the product property

$$\Vert K_{1}(X_{1},\phi)K_{2}(X_{2},\phi)\Vert_h \le \Vert
K_{1}(X_{1},\phi)\Vert_h \Vert K_{2}(X_{2},\phi)\Vert_h. 
$$  

\\which was the basis for proofs in [BDH-est].  We assume as earlier
that the activity K is supported on connected polymers.  We then
define

$$\Vert K\Vert_{h,G_{\k},\AA}=\sup_{\Delta}\sum_{X\supset\Delta}
\Vert (K(X)\Vert_{h,G_{\k}}\AA (X) \Eq(2.11)
$$

\\In addition we define {\it kernel norms}:

$$\vert K(X)\vert_{h'} =\sum_{j=0}^{n_0} {h'\over j!}
\Vert (D^{j}K)(X,0)\Vert \Eq(2.12)
$$

\\where $h'$ is a real parameter and   $h'\ge 0$. We define

$$\vert K\vert_{h',\AA}=\sup_{\Delta}\sum_{X\supset\Delta}\vert K(X)\vert_{h'} \AA (X)
\Eq(2.13)
$$

\\These definitions are also different from the kernel norms in
[BDH-est,BDH-eps]), but retain the product property. 
When using these norms for our model we will choose $n_0 =9$, $h=\e^{-1/4}$ 
and either $h'=h$ or $h'=h_{*}$, where

$$h_{*}=L^{(3-2[\phi])/2}=L^{(3+\e)/4}$$
The
kernel norms with $h'=h_{*}$ will be useful for controlling flow coefficients.
\vglue1.truecm

\\{\bf 3. RG STEP }
\numsec=3\numfor=1
\vskip.5truecm
\\In this section we describe the RG step. This consists of two 
parts: Fluctuation integration and rescaling, followed by the
extraction of relevant parts.

\vskip.5truecm \\{\bf 3.1 Fluctuation integration and rescaling}
\vskip.3truecm \\ The integration over the fluctuation field exploits
the independence of $\zeta (x)$ and $\zeta (y)$ when $|x-y|\ge L$. To
do this we pave $\Lambda$ by blocks of side $L$ called $L-$blocks so
that each $L-$block is a union of the original $1-$blocks. Let
$\bar{X}^{L}$ denote the $L-$closure of a set $X$, namely the smallest
union of closed $L-$blocks containing $X$. The polymers will be
combined into larger $L-$polymers which by definition are closed,
connected and unions of $L-$blocks.  The combination is performed in
such a way that the new polymers are associated to independent
functionals of $\zeta$.

\vskip.2truecm
\\We start with the representation (1.25) of Section 1.3.

$${\cal Z}(\Lambda)=[{\cal E}xp(\square\> e^{-V}+K)](\Lambda)  \Eq(3.1) $$

\\with $K (X) = 0$ unless $X$ is closed and connected. By definition we have:

$${\cal E}xp(\square\> e^{-V}+K)(\Lambda) = 
\sum_{N}{1\over N !}{\sum}_{(X_{j})} e^{-V(\Lambda \setminus \cup X_{j})}
{\prod}_{j=1}^{N} {K}(X_{j})   \Eq(3.4.1)$$ 

\\where the sum is over 
sequences of disjoint polymers $ (X_{j}) = ( X_{1}.....X_{N})$.
\medskip
\\Let us define: $\cup X_{j} = X $ and $\Lambda \setminus {X} = X_{c}$.
$X_{c}$ is an open set. We denote by $\bar{X}_{c}$ its closure. Obviously,
$V(X_{c}) = V(\bar{X}_{c})$, since $V(X_{c})$ is given by a 
Lebesgue integral
over $X_{c}$. Hence we can write:

$$e^{-V(X_{c} ,\zeta + \phi)} = 
{\prod}_{{\Delta}\subset\bar{X}_{c}}e^{-V(\Delta ,
\zeta + \phi)}  $$

\\Define the polymer activity P, supported on closed unit blocks, as:

$$P(\Delta ,\zeta ,\phi) =e^{-V(\Delta ,\zeta + \phi)} - e^{-\tilde{V}
(\Delta  , \phi )}  \Eq(3.5)$$

\\with $\tilde{V}$ to be chosen. In the following $V,K$ has field argument
$\zeta + \phi $ whereas $\tilde{V}$ depends only on $\phi$. The dependence
of P on $\zeta ,\phi$ is as defined above.

\medskip
\\Now write:

$$e^{-V(X_{c})}= {\prod}_{{\Delta}\subset 
\bar{X}_{c}}[e^{-\tilde{V}(\Delta)} + P(\Delta)] $$

\\then expand the product and insert the expansion into \equ(3.4.1):

$${\cal E}xp(\square\> e^{-V}+K)(\Lambda) = 
\sum {1\over N !M!}
{\sum}_{(X_{j}),(\Delta_{i})}
 e^{-\tilde{V}(X_{0})}
{\prod}_{j=1}^{N} {K}(X_{j})   {\prod}_{i=1}^{M} {P}(\Delta_{i}) \Eq(3.4)$$ 

\\where $X_{0} = \Lambda \setminus (\cup X_{j}) \cup (\cup
\Delta_{i})$. Let $Y$ be the $L-$closure of $(\cup X_{j}) \cup (\cup
\Delta_{i})$ and let $Y_{1},\dots , Y_{P}$ be the connected components
of $Y$.  These are $L-$polymers. Let $f$ be the function that maps
$\pi: = (X_{j}),(\Delta_{i})$ into $\{Y_{1},\dots ,Y_{P} \}$. Now we
perform the sum over $(X_{j}),(\Delta_{i})$ in \equ(3.4) by summing
over $\pi \in f^{-1} (\{Y_{1},\dots ,Y_{P} \})$ and then
$\{Y_{1},\dots ,Y_{P} \}$.  The result is:

$$
{\cal E}xp(\square\> e^{-V}+K)(\Lambda) = {\cal E}xp_{L}(\square\>
e^{-\tilde{V}}+ {\cal B} K)(\Lambda) \Eq(3.12)
$$

\\where the subscript on ${\cal E}xp_{L}$ indicates that that the
domain is functionals of $L-$polymers and

$$({\cal B}K)(Y) =  \sum_{N+M\ge 1}{1\over N !M!}
{\sum}_{(X_{j}),(\Delta_{i})\rightarrow \{Y \}} 
e^{-\tilde{V}(X_{0})}
{\prod}_{j=1}^{N}K(X_{j})
{\prod}_{i=1}^{M} {P}(\Delta_{i})   \Eq(3.13)$$

$$
X_{0}=Y\setminus (\cup X_{j})\cup (\cup\Delta_{i})
$$

\\where the $\rightarrow $ is the map $f$. This representation
\equ(3.12) is a sum over products of polymer activities $({\cal B}\hat
{K})(Z_{j})$ where the closed disjoint polymers $Z_{j}$ are
necessarily separated by a distance $\ge L$ and the spaces between the
polymers are filled with $ e^{-\tilde{V}}$ which are independent of
the fluctuation field $\zeta$.  The covariance $\Gamma (x-y) = 0$ if $
\vert x-y \vert \ge L$. So the fluctuation integral factorises and we
obtain:

$$\int{d\mu} _{\Gamma}(\zeta)
[{\cal E}xp(\square\> e^{-V}+K)](\Lambda ,\zeta + \phi) =
{\cal E}xp_{L}(\square_{L}\> e^{-\tilde{V}}+ 
({\cal B}K^{\sharp})(\Lambda,\phi) \Eq(3.14)$$ 

\\where the superscript $\sharp$ (``sharp'') denotes integration with
the measure $d\mu_{\Gamma}(\zeta)$.

\\Now we perform the rescaling. This is accomplished by replacing $\phi$
by the rescaled field $\RR \phi$ where

$$ (\RR\phi)(x)= \phi_{L^{-1}}(x)= L^{-[\phi]}\phi(x/L)$$
$$(\RR K)(L^{-1}X,\phi)= K(X,\phi_{L^{-1}})$$

\\If we now define $\SS =\RR{\cal B}$ then we have:

$$\int{d\mu} _{\Gamma}(\zeta)
[{\cal E}xp(\square\> e^{-V}+K)](\Lambda ,\zeta + \phi_{L^{-1}}
) =
{\cal E}xp(\square \> e^{-\tilde{V_{L}}}+ 
(\SS K)^{\natural})(L^{-1}\Lambda,\phi) \Eq(3.15)$$

\\where $\tilde{V_{L}}(\Delta , \phi)= \tilde{V}(L\Delta
,\phi_{L^{-1}})$ and the superscript $\natural$ (``natural'') denotes
integration with the measure $d\mu_{\Gamma_{L}}(\zeta)$. Here:

$$\Gamma_{L}(x-y)= L^{2 [\phi]}\Gamma(L(x-y))$$
\vskip.5truecm

\\We have returned to a functional of the form ${\cal E}xp(\square\>
e^{-V}+K)(\Lambda)$ with $V \rightarrow \tilde{V}$ and $K\rightarrow
(\SS K)^{\natural}$.  Thus the operation is an evolution of the
interaction described in coordinates $V,K$.

\vskip.2truecm
\\We will refer in the future to \equ(3.15) as the 
{\it {fluctuation step.}} The RG step will be completed by removing
relevant parts from $(\SS\hat {K})^{\natural}$ and compensating by
a new local potential, this operation being called 
{\it Extraction}.


\vskip.5truecm
\\{\bf 3.2 Extraction }
\medskip

\\The objective is to cancel parts $Fe^{-V}$ of $K$ in ${\cal
E}xp(\square\> e^{-V}+K)(\Lambda)$ by a change in $V$, adding terms
$V_{F}$ to $V$.  This will be possible for functionals $F$ of a
special form which we first describe. 

\vskip.2truecm
\\Let $ {\cal P}$ and $ {\cal P}_{j}$ be polynomials.  For $Y$, any union of cells,
$V$ has the form:
$$V(Y)=\int_{Y} dx\ {\cal P}(\phi (x))$$ 

\\We define polymer activities

$$F(X)=\sum_{\Delta \subset X}F(X,\Delta)  \Eq(33.2)$$

\\where $F(X,\Delta)$ has the form

$$F(X,\Delta)=\sum_j 
\int_{\Delta}dx \  \alpha_j (X,\Delta,x)
{{\cal P}_j}(\phi (x)) \Eq(33.2a)
$$

\\and $F(X,\Delta)=0$ for $\Delta \not \subset \Lambda $. We also define:

$$V_{F}(\Delta) = \sum_{X\supset \Delta}F(X,\Delta)  \Eq(33.2b)$$

\\Then

$$V_{F}(\Delta)=\sum_j  \int_{\Delta}dx \  \alpha_j (\Delta,x){{\cal
P}_j}(\phi (x))
=\sum_j  \int dx \  \alpha_j (x) {{\cal
P}_j}(\phi (x))
$$

\\where $\alpha (x) = \alpha (\Delta,x)$ with $\Delta \ni x$ and

$$\alpha_j (\Delta,x)  = \sum_{X\supset \Delta}\alpha_j (X,\Delta,x) \Eq(33.2c)$$

\\and for any polymer $X$ 

$$V_{F}(X)=\sum_{\Delta \subset X}V_{F}(\Delta) 
 =\sum_j \int_{X}dx \  \alpha_j (x) {{\cal P}_j}(\phi (x)) \Eq(33.2d)    $$

\\We define $V_{F}(X_{c})$ by the last member of \equ(33.2d), replacing
$X$ by $X_{c}$.

\vskip.2truecm
\\Following [BDH-est] Section~4.2, 

\vskip .5cm
\\{\bf  Extraction Formula:}

$${\cal E}xp(\square\> e^{-V}+K)(\Lambda)={\cal E}xp(\square\> e^{-V'}
+{\cal E}(K,F)](\Lambda)   \Eq(33.15)$$

\\where

$$
\eqalign{
&V'= V - V_{F}\cr
&{\cal E}_{1}(K,F) = K-Fe^{-V}\cr
}\Eq(33.12)
$$

\\where ${\cal E}_{1}(K,F)$ is the linearization of ${\cal E}(K,F)$ in
$K$ and $F$.

\vskip.2truecm 
\\The complete formula for ${\cal E}(K,F)$ is described
at the end of this section, but in fact we will only need a crude
estimate on the nonlinear part of it which will be quoted from
[BDH-est].

\vskip.2truecm
\\We will need a variant of the extraction formula in which vacuum
energy is factored out completely as follows. Suppose $F$ has an
additive piece  $F_0$ which is  field independent, i.e. if it is of
the form 
$$F=F_1+F_0\Eq(33.16)$$
we have
$$V'_F=V-V_F=V'_{F_1}-V_{F_0}$$
and then
$${\cal E}xp(\square\> e^{-V'_F}
+{\cal E}(K,F)(\Lambda)=e^{V_{F_0}(\L)}{\cal E}xp(\square\> e^{-V'_{F_1}}
+{\cal E}(K,F_0,F_1)(\Lambda)\Eq(33.17)$$
where
$${\cal E}(K,F_0,F_1)=e^{-V_{F_0}}[{\cal E}(K,F)]\Eq(33.18)$$

\\The extraction formula \equ({33.17}) is applied to \equ({3.15}) so
that the combination of integrating out the fluctuation field and
extracting returns ${\cal E}xp(\square\> e^{-V}+K)(\Lambda)$ to a
functional of the same form with $V \rightarrow \tilde{V}'_{F_{1}}$
and $K\rightarrow {\cal E}((\SS K)^{\natural},F_0,F_1)$. This is a
complete RG step .  

\vskip .5cm
\\{\bf Formulas for ${\cal E} (K,F)$:}
\vskip .2cm
\\$${\cal E}(K,F)(W)=\sum_{M,N}{1\over M!}{1\over N!}
\sum_{(Z_{j}),(Y_{k})\rightarrow W} \ e^{-V'(W\setminus Y)} $$
$$\prod_{j=1}^{M}(e^{-F(Z_{j},Z_{j}\cap \bar Y_{c})}-1)\prod_{k=1}^{N}
\tilde {K} (Y_{k}) \Eq(33.14)$$

\\where $N\ge 1,\ \ M\ge 0$ and $Y=\cup_{k=1}^{N}Y_{k}$, the $Y_{k}$ 
are disjoint and:
\medskip
\\1) the polymers $(Z_{j}),(Y_{k})$ are connected and
such that $W= (\cup Z_{j})\cup (\cup Y_{k})$
\medskip
\\2) for every $j$, $Z_{j}\not \subset Y_{c}, Z_{j}\not \subset Y$
\medskip
\\and

$$F(Z,Z\cap \bar Y_{c}) =\sum_{\Delta \subset Z\cap \bar Y_{c}}F(Z,\Delta) 
\Eq(33.10)$$

$$\tilde K(X)=K(X)-e^{-V(X)}(e^{F}-1)^{+}(X) \Eq(33.4) $$

$$J^{+} (X) = \sum_{N>1}{1\over N!} \sum_{(X_{j})\rightarrow X}\prod J (X_{j})$$

\\where $(X_{j})\rightarrow X$ iff $X = \cup X_{j}$ and $X_{j}$ are
distinct (as opposed to disjoint) sets. $J (X) = e^{F (X)}-1$. $J^{+}$
is a polymer activity (vanishes when $X$ is not closed and connected).

\vskip.3truecm 
\\{\it Remark}: The proof of \equ({33.14}) is step by step the same
as Sec. 4.2 of [BDH-est] with appropriate changes to take into account
the use of closed polymers instead of open polymers.

\vskip.3truecm 
\\{\it Appendix}: For convenient later reference we collect here the
notations associated with rescaling that will be used in this
paper.  Some of them refer to objects not yet introduced.

$$\RR\phi(x) =\phi_{L^{-1}}(x) =L^{-[\phi]}\phi(x/L)\Eq(1.18.2)$$

$$(\RR K)(L^{-1}X,\phi)= K_{L}(L^{-1}X,\phi)= K(X,\phi_{L^{-1}})$$

$$V_{L}(\Delta , \phi)= V(L\Delta ,\phi_{L^{-1}})$$

$$\Gamma_{L}(x-y)= L^{2 [\phi]}\Gamma(L(x-y))$$

\\For an integral kernel $u(x-y)$, e.g. a covariance, we define
a rescaled version

$$u_L(x-y)=L^{2[\phi]}u(L(x-y))=
L^{3-\e\over 2}u(L(x-y))\Eq(4.3)$$

$$\eqalign{\tilde V_{L}(\Delta , g,\m)=&{V}(\Delta ,\phi,C,g_L,\m_L)
\cr
g_L=L^\e g,&\quad\m_L=L^{3+\e\over 2}\m\cr}\Eq(4.6)$$

\\A set $X$ is said to be {\it small} if $X$ is connected and $|X|\le
2^{D}$. In the present case the dimension $D=3$. $\bar{X}^{L}$ is the
$L$-closure of $X$, which by definition is  smallest union of closed
$L$-blocks containing $X$.

\vskip.3truecm 
\\The superscript $\natural$ (``natural'') denotes integration with
the measure $d\mu_{\Gamma_{L}}(\zeta)$ and the superscript $\sharp$
(``sharp'') denotes integration with the measure
$d\mu_{\Gamma}(\zeta)$.

\vskip.5truecm
\\{\bf 4. APPLICATION OF RENORMALIZATION GROUP STEP  }
\numsec=4\numfor=1
\vskip.5truecm

\\In this section we specify a particular RG map by making choices for
$\tilde{V}, F$ in \equ({3.5}) and \equ({33.12}). We define the second
order approximation to the RG map and derive formula for the error
after second order. 

\\Recall:

$$V(\D,\phi)=V(\D,\phi,C,g,\m)=g\int_\D d^3x:\phi^4:_C(x)+
\m\int_\D d^3x:\phi^2:_C(x)\Eq(4.1)$$

\\We define $\tilde V$ by

$${\tilde V}(\D,\phi)=V(\D,\phi,C_{L^{-1}},g,\m)=g\int_\D 
d^3x:\phi^4:_{C_{L^{-1}}}(x)+
\m\int_\D d^3x:\phi^2:_{C_{L^{-1}}}(x)\Eq(4.2)$$

\vskip.5truecm
\\Recall that the RG acts on functionals written in the form ${\cal
E}xp(\square\> e^{-V}+K)(\Lambda)$.  In this section we refine this
description by writing

$$K=Qe^{-V}+R 	\Eq(4.12.5)$$

\\where $Q$ is an explicit polymer activity which we will call the
{\it``second order polymer activity''}. It will be derived from second
order perturbation theory in powers of $g$ and is defined as follows:
$Q$ is supported on connected polymers $X$, $|X|\le 2$. We write

$$Q(X,\phi)=Q(X,\phi;C,{\bf w},g) = 
g^{2}\sum_{j=1}^{3}n_{j}Q^{(j,j)}(\tilde X,\phi;C,w^{(4-j)})\Eq(4.19.1)$$

\\where $(n_{1},n_{2},n_{3})= (48,36,8)$ and ${\bf
w}=(w^{(1)},w^{(2)},w^{(3)})$ is a triple of integral kernels to be
obtained inductively and

$$\tilde X=\cases{\D\times\D&if $X=\D$\cr
(\D_1\times\D_2)\cup(\D_2\times\D_1)&if $X=\D_1\cup\D_2$\cr
0&otherwise}\Eq(4.17)
$$

$$\eqalign{Q^{(m,m)}(\tilde X,\phi;C,w^{(4-m)})=&{1\over 2}
\int_{\tilde X}\!\!d^3xd^3y:\!(\phi^m(x)\!-\!\phi^m(y))^2\!:_C
w^{(4-m)}(x-y)\quad {\rm for}\ m=1,2\cr
Q^{(3,3)}(\tilde X,\phi;C,w^{(1)})=&{1\over 2}
\int_{\tilde X}d^3xd^3y:\phi^3(x)\phi^3(y):_C
w^{(1)}(x-y)\cr}\Eq(4.18)$$

\vskip.5truecm
\\Next we define the second order approximation to the RG map.
We say that an activity $p (X)$ is supported on (closed/open) unit
blocks if $p (X)=0$ for $X$ not a (closed/open) unit block. A block is
closed by default. Let $p$ be the activity supported on unit blocks
defined by

$$p(\D,\z,\phi)=V(\Delta ,\zeta + \phi)-\tilde{V}
(\Delta  , \phi ) 
 = p_{g} + p_{\mu}
\Eq(4.8a)$$

\\where 

$$p_{g}=g\int_\D d^3x\left(:\z^4:_\G(x)+4\phi(x):\z^3:_\G(x)
+6:\phi^2:_{C_{L^{-1}}}(x):\z^2:_\G(x)+\right.$$
$$\left. 4:\phi^3:_{C_{L^{-1}}}(x)\z(x)\right)
$$

$$p_{\mu}=\m\int_\D d^3x\left(2\phi(x)\z(x)+:\z^2:_\G(x)\right)
\Eq(4.8b)$$

\\We insert a parameter $\lambda$ into our previous definitions in
such a way that (i) at $\lambda =1$ our $\lambda$ dependent objects
correspond with the previous definitions. (ii) The expansion through
order $\lambda^{2}$ is second order perturbation theory in $g$
counting $\mu =O (g^{2})$. (iii) Powers of $\lambda$ are determined so
as to correspond with leading powers of $g$ buried inside polymer
activities. (iv) All functions will turn out to be norm analytic in
$\lambda$. Thus

$$
P (\lambda) =  e^{-\tilde{V}} \big(
-\lambda p_{g}-\lambda^{2}p_{\mu}+{1\over 2}\lambda^{2}p_{g}^{2}
\big) + \lambda^{3}r_{1}
\Eq(4.9)$$

\\where $r_{1}$ is defined by the condition $P (\lambda = 1) = P =
e^{-V} - e^{-\tilde{V}}$.  Similarly, we define

$$
K (\lambda) = 
\lambda^{2}e^{-\tilde{V}}Q  +\lambda^{3} \big(
[e^{-V} - e^{-\tilde{V}}]Q + R 
\big)
\Eq(4.10)$$

\\which, for $\lambda =1$ coincides with $K = e^{-V}Q+R$, where $Q$
will be an explicit polymer activity obtained by second order
perturbative calculation and $R$ will be the remainder after second
order perturbation theory.  Corresponding to \equ({3.13}) we define

$${\cal B}(\lambda ,K) (Y) =  \sum_{N+M\ge 1}{1\over N !M!}
{\sum}_{(X_{j}),(\Delta_{i})\rightarrow \{Y \}} 
e^{-\tilde{V}(X_{0})}
{\prod}_{j=1}^{N} K (\lambda ,X_{j}) 
{\prod}_{i=1}^{M}P (\lambda ,\Delta_{i})   \Eq(4.13)$$

$$
X_{0}=Y\setminus (\cup X_{j})\cup (\cup\Delta_{i}) \Eq(4.11)
$$

\\Let $\SS(\lambda,K) = \RR\circ {\cal B} (\lambda,K)$,
where $\RR$ is the rescaling defined in the last section.  The RG
evolution for $K$ with parameter $\lambda$ is $f_{K}: K\mapsto {\cal
E} (\SS (\lambda,K)^{\natural},F (\lambda ))$, where

$$F (\lambda) = \lambda^{2}F_{Q}+ \lambda^{3}F_{R}   \Eq(4.12)$$

\\will be specified later, and $\natural$ is the rescaled integration
over $\zeta$, and, as usual, $F (\lambda)= F$, when $\lambda =1$.
Given a function $f (\lambda)$ let

$$T_{\lambda}f = f (0)+f' (0)+{1\over2}f'' (0)	\Eq(4.13.5)$$ 

\\be the Taylor expansion to second order evaluated at $\lambda
=1$. Then the second order approximation to the RG map is  $f^{(\le 2)} =
(f^{(\le 2)}_{K},f^{(\le 2)}_{V})$ with

$$
f^{(\le 2)}_{K} (K,V) 
= T_{\lambda} {\cal E} (\SS(\lambda,K)^{\natural},F (\lambda ))
= {\cal E}_{1} ( T_{\lambda} \SS(\lambda,K)^{\natural},F_{Q})
\Eq(4.14)$$

$$f^{(\le 2)}_{V}(K,V) = V'_{F} 	\Eq(4.14.1)$$

\\Note also that only the linearized ${\cal E}_{1}$ intervenes, because
it will turn out that the nonlinear part of extraction generates terms
only at order $\lambda ^{3}$ or higher.

\vskip.5truecm
\\{\it Proposition 4.1
\vskip.2truecm
\\There is a choice of $F_{Q}$ such that
the form of $Q$ remains invariant under the RG evolution at second
order. In more detail, $f^{(\le 2)}(V,Qe^{-V}) =
(V'_{F_{Q},(\le 2)},Q'_{(\le 2)}e^{-\tilde V_{L}})$ where the parameters in
$V'_{(\le 2)}$ evolved according to

$$\eqalign{
&V'_{(\le 2)}(\D)=V(\D,C,g'_{(\le 2)},\m'_{(\le 2)})\cr
&g'_{(\le 2)}=L^\e g(1-L^\e ag)\cr
&\m'_{(\le 2)}=L^{3+\e\over 2}\m-L^{2\e}bg^2\cr}
\Eq(4.47)$$

\\The parameters in $Q'_{(\le 2)}$ evolved according to

$$
\eqalign{
&Q'_{(\le 2)}=Q(C,{\bf w}',g_{L})\cr
&{\bf w}'={\bf v}+{\bf w}_L\cr
&v^{(1)}=\G_L, \qquad 
v^{(p)}=(C_L)^p-(C)^p\qquad 2\le p\le 4\cr
}    \Eq(4.54)
$$
}

\vskip.5truecm
\\{\it Proof:} We define a polymer activity $\hat Q$ supported on
connected polymers $X$ with $|X|\le 2$

$$\hat Q(X,\z,\phi)=\cases{
{1\over 2}(p_{\lambda }(\D,\z,\phi))^2 & $|X|=\Delta$ \cr \cr
{1\over 2}\displaystyle\sum_{\D_1,\D_2\atop\D_1\cup\D_2=X}
p_{\lambda}(\D_1,\z,\phi)p_{\lambda}(\D_2,\z,\phi)
&$|X|=2$, connected\cr
}\Eq(4.9.1)$$

\\It is easy to check that

$$T_{\lambda }\SS (K,\lambda ) = 
- p_{L}e^{-\tilde{V}_{L}} + \bigg\{ 
e^{-\tilde{V}_L}\hat Q_L
+e^{-\tilde{V}_L}Q_L  
\bigg\}   \Eq(4.21.1)$$

\\Let

$$\tilde Q(C,{\bf v},g_L)= g_{L}^{2}\sum_{j=1}^{4}n_{j}\tilde Q^{(j,j)}(\tilde
X,\phi;C,v^{(4-j)})
\Eq(4.14.2)$$

\\where $(n_{1},n_{2},n_{3},n_{4})= (48,36,8,12)$  and 

$$\tilde Q^{(m,n)}(\tilde X,\phi;C,u)={1\over 2}
\int_{\tilde X}d^3xd^3y:\phi^m(x)\phi^n(y):_C
u(x-y)\Eq(4.25)$$

\\We then have

$$T_{\lambda }\SS (K,\lambda )^{\natural} =
e^{-\tilde{V}_L}\left(\tilde Q(C,{\bf v},g_L)
+Q(C,{\bf w}_L,g_L)\right)
\Eq(4.30)$$


\vskip.2truecm
\\Define

$$F_{Q} = 
\tilde Q(C,{\bf v},g_L) - Q(C,{\bf v},g_L) \Eq(4.52)
$$

\\Then we have from \equ({4.30}) and \equ({4.52})

$$
{\cal E}_{1} \bigg(T_{\lambda }\SS
(\lambda,K)^{\natural},F\bigg)
= T_{\lambda }\SS(\lambda,K)^{\natural}
-
F_{Q}e^{-\tilde{V}_{L}}
=
e^{-\tilde{V}_{L}}Q(C,{\bf w}^{(\le 2)},g_L) 
\Eq(4.15)$$

\\We write
$$F_{Q}(X,\phi)=F_{1,Q}(X,\phi)+F_{0,Q}(X)
\Eq(4.32)$$
where $F_{0,Q}$ is the field independent part of $F_{Q}$. Then
$$\eqalign{F_{1,Q}=&g_L^2\left\{36\ \tilde Q^{(4,0)}(C,v^{(2)})+
48\ \tilde Q^{(2,0)}(C,v^{(3)})\right\}\cr
F_{0,Q}=&12g_L^2\tilde Q^{(0,0)}(C,v^{(4)})\cr}\Eq(4.33)$$

\\$
F_{1,Q}(X)$ can be written as:
$$F_{1,Q}(X)=\sum_{\D\subset X}F_{1,Q}(X,\D)\Eq(4.34)$$
where
$$
F_{1,Q}(X,\D)=36g_L^2F^{(4)}_{1,Q}(X,\D)
+48g_L^2F^{(2)}_{1,Q}(X,\D)\Eq(4.35)$$
and
$$F^{(m)}_{1,Q}(X,\D)=\int_\D d^3x:\phi^m:_C(x)
f^{(m)}_{1,Q}(x,X,\D)
\Eq(4.36)$$
with
$$
f^{(m)}_{1,Q}(x,X,\D)=\cases{\int_\D d^3y v^{(m')}(x-y)&$X=\D$\cr
\cr
\int_{\D'}d^3y v^{(m')}(x-y)&$X=\D\cup\D'$, connected\cr}\Eq(4.36a)$$
and $$m' = 4- m/2$$ 

$$V(F_{1,Q},\D)=
\sum_{X\supset\D}F_{1,Q}(X,\D)=36g_L^2\sum_{X\supset\D}
F^{(4)}_{1,Q}(X,\D)
+48g_L^2\sum_{X\supset\D}F^{(2)}_{1,Q}(X,\D)\Eq(4.37)$$
In the following we will use the:
\vglue.3truecm
\\{\it Claim:}

\\$v^{(j)},\ 1\le j\le 4$ are $C^\io$, positive, and have support
$v^{(j)}(x-y)=0$ for $|x-y|\ge 1$.

\vglue.3truecm
\\{\it Proof:}
That they are
$C^\io$ follows from their definition \equ(4.15), $\G$ is  $C^\io$ 
and $C$ is  $C^\io$ .
For the support property, this is obvious for $v^{(1)}=\G_L$, and for
$p\ge 2$: $v^{(p)}=C_L^p-C^p$ with pointwise multiplication. The latter
has $\G_L$ as a factor because $C_L=C+\G_L$ and $\G_L$ has the required
support property. The positivity follows from that of $\G_L$ and $C$.
This proves the claim.
\vglue.3truecm
\\Now return to \equ(4.37).
$$\sum_{X\supset\D}F^{(m)}_{1,Q}(X,\D)=
\int_\D d^3x:\phi^m:_C(x)
\left[\int_\D d^3y v^{(m')}(x-y)+\!\!\!\!\!\!\!\!\!
\sum_{\D'\ne\D\atop(\D,\D')\ {\rm connected}}\!\!\!\!\!\!
\int_{\D'} d^3y v^{(m')}(x-y)\right]$$
On the r.h.s. use $v^{(m')}(x-y)=0$ for $|x-y|\ge 1$ to extend the sum
on $\D'$ to {\it all} the $\D'\ne\D$. We then get
$$\sum_{X\supset\D}F^{(m)}_{1,Q}(X,\D)=
\int_\D d^3x:\phi^m:_C(x)\left[\int d^3y v^{(m')}(x-y)\right]$$
Hence from \equ(4.37) and above we get
$$V(F_{1,Q},\D)=ag_L^2\int_\D d^3x:\phi^4:_C(x)+b
g_L^2\int_\D d^3x:\phi^2:_C(x)\Eq(4.38)$$
where
$$\eqalign{a=&36\int d^3y\ v^{(2)}(y)>0 = O(log \ L) >0\cr
b=&48\int d^3y\ v^{(3)}(y)>0 =O(L^{3/2})>0   \cr}\Eq(4.39)$$
That $a$ and $b$ are well defined and positive is easy to see
using the claim above. That 
$a=O(log \ L),b=O(L^{3/2})$ is proved later
in Section 5 ( see Lemma~5.12). End of proof.

\vskip.5truecm
\\{\it The exact RG evolution for $K=Qe^{-V}+R$}.
\vskip.2truecm

\\The exact map
$$
K\mapsto K' 
= f_{K} (\lambda,K,V)\vert_{\lambda =1} 
=  {\cal E} (\SS (\lambda,K)^{\natural},F (\lambda
))\vert_{\lambda =1}
\Eq(4.16)$$

\\induces an evolution of the remainder $R$ which is studied by Taylor
series around $\lambda =0$ with remainder written using the Cauchy
formula:

$$
f_{K} (\lambda =1) = \sum_{j=0}^{3} {f_{K}^{(j)} (0)\over j!} + 
{1\over 2\pi i} 
\oint_{\gamma} {d\lambda \over \lambda^{4} (\lambda -1)}
f_{K} (\lambda)
$$

\\The terms $j=0,1,2$ are the second order part $f^{(\le 2)}_{K}$.  In
the $j=3$ term there are no terms mixing $R$ with $Q,P$ because of the
$\lambda^{3}$ in front of $R$.  Therefore it splits

$$
{f_{K}^{(3)} (0)\over 3!} = R_{1} + R_{2} 
$$

\\into the third order derivative at $R=0$, which we write
using the Cauchy formula as

$$
R_{1} \equiv R_{\rm main} 
= {1\over 2\pi i} 
\oint_{\gamma} {d\lambda \over \lambda^{4}}
{\cal E} \bigg( 
\SS(\lambda ,Qe^{-V})^{\natural},F_{Q} (\lambda ) 
\bigg)
\Eq(4.188)$$

\\and terms linear in $R$:

$$\eqalign{
&R_{2} \equiv R_{\rm linear}
= \big(\SS_1 R\big)^{\natural} - F_{R}e^{-\tilde V_{L}}\cr
&\SS_1 R(Z) = \sum_{X:L^{-1}\bar{X}^{L}=Z} e^{-\tilde V_L(Z\backslash
L^{-1}X)} R_{L}(L^{-1}X)\cr
} 	\Eq(4.19)$$

\\The remainder term in the Taylor expansion is

$$
R_{3} = 
{1\over 2\pi i} 
\oint_{\gamma} {d\lambda \over \lambda^{4} (\lambda -1)}
{\cal E} (\SS (\lambda,K)^{\natural},F (\lambda
))
\Eq(4.21)$$

\\In Proposition~4.1 the coupling constant in $Q$ is not the same as
the coupling constant in $V^{(\le 2)}$.  Furthermore, the coupling
constant in $V^{(\le 2)}$ will further change because of the
contribution from $F_{R}$.  To take this into account we introduce

$$\eqalign{
&V'(\D)=V(\D,C,g',\m')\cr
&g'=L^\e g(1-L^\e ag)+\xi_{ R}\cr
&\m'=L^{3+\e\over 2}\m-L^{2\e}bg^2+\r_{ R}\cr}
\Eq(4.47A)$$

\\where the remainders $\xi_{ R}, \r_{ R}$ anticipate the effects of a
yet-to-be-specified $F_{R}$.  Then we set

$$
R_{4} = e^{-V'}Q(C,{\bf w}',g')- e^{-\tilde{V}_{L}}Q(C,{\bf w}',g_L)
\Eq(4.22)
$$

\\With these definitions we have written the RG as a map

$$
{\cal E}xp(\square\> e^{-V}+Qe^{-V}+R)(\Lambda) \mapsto 
{\cal E}xp(\square\> e^{-V'}+Q'e^{-V'}+R')(L^{-1}\Lambda)
$$

\\with $Q'= Q(C,{\bf w}',g')$ and the RG map $f_{K}$ has induced a map
$f_{R}$

$$
R' \equiv f_{R}(V,R) = R_{\rm main} + R_{\rm linear} + R_{3} + R_{4}
\Eq(4.23)$$

\vskip.5truecm
\\{\it Construction of $F_{R}$}
\vglue.3truecm

\\To complete the RG step we must specify the relevant part from the
remainder $F_{R}$.   The goal is to choose $F_{R}$ so that the map $R
\rightarrow R_{\rm linear}$ will be contractive.  As we will later
prove, this will be the case provided the small set part of $R_{\rm
linear}$ is normalized so that certain derivatives with respect to
$\phi$ vanish at $\phi =0$.  We say that a functional $J$ is
normalized if, for all small sets $X$,

$$\eqalign{&J(X,0)=0\cr
&D^2J(X,0;1,1)=0\cr
&D^2J(X,0;1,x_\m)=0\cr
&D^4J(X,0;1,1,1,1)=0\cr}\Eq(4.61)
$$

\\Define

$$\tilde F_{R}(X,\phi)= \sum_{P} \int_Xd^3x\ 
{\tilde \alpha}_{P} (X)P (\phi (x),\dpr \phi (x))
\Eq(4.57)
$$


\\where $P$ runs over the {\it relevant} monomials, which , in this
model are

$$
P = 1,
\ \phi^{2},
\ \phi^{4},
\ \phi \dpr_{\m}\phi  \ {\rm with } \ \mu =1,2,3,
$$

\\Choose the coefficients ${\tilde \alpha}_{P}$ so that

$$J= R^{\sharp}-\tilde F_{R}e^{-\tilde{V}}
\Eq(4.59)
$$

\\is normalized (details are given below). We define the relevant
part, a functional supported on small sets, by

$$F_{R}(Z)=\sum_{X: {\rm small\ sets}\atop L^{-1}\bar{X}^{L}=Z}
\tilde F_{R,L}(L^{-1}X) 	\Eq(4.58)$$

\\Then, from the definition of $R_{\rm linear}$ in \equ(4.19),

$$
R_{\rm linear} (Z)
= \sum_{X: {\rm small\ sets}\atop L^{-1}\bar{X}^{L}=Z}
e^{-\tilde{V}_{L} (Z\setminus L^{-1}X)}J_{L} (L^{-1}X)
+
\sum_{X: {\rm large\ sets}\atop L^{-1}\bar{X}^{L}=Z}
e^{-\tilde{V}_{L} (Z\setminus L^{-1}X)}J_{L} (L^{-1}X)
\Eq(4.59b)$$

\\Therefore the first sum in $R_{\rm linear}$ is also normalized
because normalization is preserved under multiplication by smooth
functionals of $\phi$ and rescaling.

\vskip.3truecm
\\Having defined $F_{R}$, we must show that it has the form
\equ({33.2}), \equ({33.2a}) required for the extraction operation. 
Define $\tilde{F}_{R} (X,Y)$ by replacing the integral over $X$ by
integration over $X\cap Y$ in \equ({4.57}). Let

$$F_{R}(Z,\D,\phi)=
\sum_{X\ {\rm small\ set}\atop L^{-1}\bar{X}^{L}=Z}
\tilde F_{R}(X,L\D\cap X,\phi_{L^{-1}})
\Eq(4.67)$$

\\It is clear that \equ({33.2}) holds.  For each monomial $\alpha_{P}
(X) P$ in \equ({4.57}) define

$$
\alpha_{P}(Z,\Delta,x) = 
\sum_{X\ {\rm small\ set}\atop L^{-1}\bar{X}^{L}=Z}
{\tilde \alpha}_{P}(X) L^{-[P]+3}1_{\Delta \cap L^{-1}X} (x)
\Eq(4.67a)$$

\\where $[P]$ is the dimension of the monomial $P$, ($n[\phi]$ for
$\phi^{n}$ and $2[\phi]+1$ for $\phi \dpr \phi $).  Then

$$F_{R}(Z,\D,\phi)= \sum_{P}\int d^3x\ 
\alpha_{P} (Z,\Delta ,x)P (\phi (x),\dpr\phi (x))
\Eq(4.57b)
$$

\\which shows that \equ({33.2a}) holds.  

\vskip.2truecm
\\In order to compute $V_{F_{R}}$, note that by translation
invariance,

$$\eqalign{
\sum_{Z\supset \Delta}\alpha_{P}(Z,\Delta ,x) &= \sum_{Z\supset \Delta}
\sum_{X\ {\rm small\ set}\atop L^{-1}\bar{X}^{L}=Z}
\alpha_{P}(X) L^{-[P]+3}1_{\Delta \cap L^{-1}X} (x)\cr
&= \sum_{X\ {\rm small\ set}\atop L^{-1}\bar{X}^{L}\supset \Delta}
\alpha_{P}(X) L^{-[P]+3}1_{\Delta \cap L^{-1}X} (x)\cr
&= \sum_{X\ {\rm small\ set}}
\alpha_{P}(X) L^{-[P]+3}1_{\Delta \cap L^{-1}X} (x)\cr
&=: \alpha_{P} \qquad {\rm a.e}\cr
}
\Eq(4.57c)$$

\\is independent of $x,\Delta$, (equality a.s. because the equality
fails for $x$ in boundaries of blocks).  This can be written more
simply as

$$
\alpha_{P} = L^{-[P]+3} 
\sum_{X\ {\rm small\ set} \supset \Delta (x)} \alpha_{P}(X)
\Eq({4.70})$$

\\where $\Delta (x)$ is the block that contains $x$, excluding $x$ in a
boundary.  Therefore, with coefficients derived as in \equ({4.70}),

$$V(F_{R},\D)=\sum_{Z\supset\D}F_{R}(Z,\D)=
\sum_{P}\alpha_{P}\int_{\D}
d^3x \,P (\phi)$$
$$=:\int_{\D} d^3x\bigg\{
\alpha _{0} + \alpha_{2,0} \phi^2 +
\alpha_{4,0} \phi^{4}
\bigg\}
\Eq(4.68)
$$

\\where the term $\a_{2,1} \phi(x)\dpr_\m\phi(x)$ is absent because
$\a_{2,1} =0$ by reflection invariance.  We have to rewrite this in a
$C$ Wick ordered basis in order to compute $V'$

$$
V(F_{R},\D)= \int_\D d^3x
\bigg\{\beta _{0}+
\r_{R}:\phi^2:_C(x)+
\xi_{R}:\phi^4:_C(x)
\bigg\}
\Eq(4.71)
$$

\\where

$$\eqalign{
\b_0=&\a_0+C(0)\a_{2,0}+3C^2(0)\a_{4,0}\cr
\r_{R}=&\a_{2,0}+6C(0)\a_{4,0}\cr
\xi_{R}=&\alpha_{4,0}}
\Eq(4.64)$$

\\which are formulas for the error terms in \equ(4.47).

\vskip.5truecm
\\ {\it Determining coefficients from \equ({4.61})}
\vskip.3truecm

\\
  Note that the odd derivatives $D^jJ(X,0;f^{\times j})$, $j$=odd integer,
vanish identically by $\phi\rightleftharpoons -\phi$ symmetry.
Taking derivatives of \equ(4.59) we get
$$\eqalign{J(X,0)&=R^{\sharp}(X,0)-\tilde F_{R}(X,0)
e^{-\tilde{V}(X,0)}\cr
D^2J(X,0;f^{\times 2})&=D^2R^{\sharp}(X,0;f^{\times 2})-
D^2\tilde F_{R}(X,0;f^{\times 2})e^{-\tilde{V}(X,0)}+\cr
&+\tilde F_{R}(X,0)D^2\tilde V(X,0;f^{\times 2})
e^{-\tilde{V}(X,0)}\cr
D^4J(X,0;f^{\times 4})&=D^4R^{\sharp}(X,0;f^{\times 4})+
D^4\tilde F_{R}(X,0;f^{\times 4})e^{-\tilde{V}(X,0)}+\cr
&+4D^2\tilde F_{R}(X,0;f^{\times 2})D^2\tilde V(X,0;f^{\times 2})
e^{-\tilde{V}(X,0)}\cr
&+\!\tilde F_{R}(X,0)D^4\tilde V(X,0;f^{\times 4})
e^{-\tilde{V}(X,0)}\!\!-\!3\tilde F_{R}(X,0)(\!D^2\tilde 
V(X,0;f^{\times 2}))^2
e^{-\tilde{V}(X,0)}
}$$
Note that from \equ(4.57)
$$\eqalign{\tilde F_{R}(X,0)=&\ 
{\tilde \a}_{0}(X)|X|\cr
D^2\tilde F_{R}(X,0;1,1)=&\ 4|X|
{\tilde \a}_{2,0}(X)\cr
D^2\tilde F_{R}(X,0;1,x_\m)=&\ |X|
{\tilde \a}_{2,1}(X,\m)\cr
D^4\tilde F_{R}(X,0;1,1,1,1)=&\ 24|X|
{\tilde \a}_{4}(X)\cr}$$

Now imposing successively the conditions \equ(4.61) we get
$$\eqalign{
{\tilde \a}_0(X)&= {1\over |X|}R^{\sharp}(X,0)e^{-\tilde{V}(X,0)}\cr
{\tilde \a}_{2,0}(X)&= {1\over 4}{1\over |X|}e^{-\tilde{V}(X,0)}
\left[D^2R^{\sharp}(X,0;1,1)+R^{\sharp}(X,0)
D^2\tilde V(X,0;1,1)\right]\cr
{\tilde \a}_{2,1}(X,\m)&={1\over |X|}e^{-\tilde{V}(X,0)}
\left[D^2R^{\sharp}(X,0;1,x_\m)+R^{\sharp}(X,0)
D^2\tilde V(X,0;1,x_\m)\right]\cr
{\tilde \a}_4(X)&= {1\over 24}{1\over |X|}e^{-\tilde{V}(X,0)}\left[
D^4R^{\sharp}(X,0;1,1,1,1)\right.\cr
&+D^2\tilde V(X,0;1,1)\left([D^2R^{\sharp}(X,0;1,1)+R^{\sharp}(X,0)
D^2\tilde V(X,0;1,1)\right)\cr
&+\left.R^{\sharp}(X,0)\left(D^4\tilde V(X,0;1,1,1,1)-3
(D^2\tilde 
V(X,0;1,1))^2\right)
\right]\cr}\Eq(4.62)$$
Note that the leading contributions to the ${\tilde \a}_{P}(X)$ are obtained
by setting ${\tilde V}=0$ in the above formulae. 
\vglue.3truecm

\vskip.3truecm
\\{\it Resume}

\\We have thus produced at the end of the RG step the promised map :
$$(V,Q,R)\rightarrow (V',Q',R')$$
$V'$ is the same as $V$ with evolved coupling 
$\ g\rightarrow g'\ \ \mu\rightarrow\mu'$ given by the flow (\equ({4.47A})), with 
$a,b$ given in \equ({4.39}) and 
$\xi_{R},\ \  \rho_{R}$ in \equ({4.64}). $Q'$ is the same as $Q$ with 
the 
change $\bf w\rightarrow \bf w'$, \equ({4.54}), and $R'$ is given by \equ({4.23}) 
with intervening quantities defined earlier. 

\vglue.5truecm




\\{\bf 5. ESTIMATES.}
\numsec=5\numfor=1
\vglue.3truecm
\\We will assume $L$ large but fixed and then $\e$ sufficiently small
depending on $L$ . $O(1)$ denotes a constant independent of $L$ and $\e$.
Constants C are independent of $\e$ but may depend on $L$. These 
constants may change from line to line. It will not be necessary to
keep track of these changes. 
\vskip.3truecm
{\it Throughout we will assume that {\bf w} at a generic step has been
obtained by successive iterations \equ(4.54) with initial ${\bf w_{0}}=0$}.

We make in this section the following 
hypothesis in terms of the norms introduced in Section 2.2. 
\vskip.3truecm
\\{\it  Hypothesis}
$$\vert g-\bar g\vert\le \e^{3/2},\ \ \vert\mu\vert\le \e^{2-\delta}  \Eq(55.1)$$
$$\Vert R\Vert_{h,G_{\k},\AA}\le \e^{3/4-\eta}  \Eq(55.2) $$
$$\vert R\vert_{h_{*},\AA} \le \e^{11/4-\eta}  \Eq(55.3)$$
where $\delta,\eta =O(1)> 0$ are very small fixed numbers, say $1/64$, and 
$h=c\e^{-1/4}$ with $c=O(1)$ a very small number. Further more we take
$h_{*}=L^{(3+\e)/4}$ and choose $n_{0}=9$. \\
Moreover $\bar g$ is the approximate fixed point in the flow of the coupling
constant $g$ obtained from the first equation in \equ(4.47A) by ignoring
the remainder $\xi_{\hat R}$. Namely,
$$\bar g ={L^{\e}-1\over L^{\e}a} =O(\e)>0   \Eq(55.4)$$
for $\e$ sufficiently small depending on $L$. We have used the estimate
$a=O(log\ \ L)>0$ which is proved below in Lemma~5.12 (independent
of the  Hypothesis).\\
Note that the  Hypothesis now implies that $g=O(\e)$ for 
$\e$ sufficiently small.

\vglue.3truecm

\\Recall the definitions of $\rho_{ R}$ and $\xi_{ R}$ from
\equ({4.64}). We will prove in this section the following 

\vglue.3truecm

\\{\it Theorem 1

\\Given the  Hypothesis above we have for $L$ sufficiently large
and then $\e$ sufficiently small

$$\vert\xi_{ R}\vert \le L^{\e}\e^{11/4-\eta} \Eq(555.3)$$

$$\vert\rho_{ R}\vert \le L^{(3+\e)/2}\e^{11/4-\eta} \Eq(555.4)$$

$$\vert g'-\bar g\vert\le \e^{3/2},
\ \ \vert\mu'\vert\le C\e^{2-\delta} 
\Eq(55.5)
$$

$$\vert g'-g\vert\le \e^2   \Eq(55.6) $$

$$\Vert R'\Vert_{h,G_{\k},\AA}\le L^{-1/4}\e^{3/4-\eta}  \Eq(55.7)$$

$$\vert R'\vert_{h_{*},\AA} \le L^{-1/4}\e^{11/4-\eta}  \Eq(55.8)$$
}

\vskip.3truecm

\\The following long series of lemmas, {\it except} Lemmas 5.1-5.4
and 5.14, are proved under the Hypothesis above, and will serve to
prove Theorem 1.  Lemmas~5.21, 5.22, 5.23 and 5.27 are the major
parts of the program. $R_{\rm main}$ is bounded in Lemma~5.21 and
this result determines the qualitative form of the bound on the
remainder.  $R_{3}$ and $R_{4}$ are seen, in Lemmas~5.22, 5.23 to be
negligible in comparison.  $R_{\rm linear}$ is the crux of the program
and it is bounded in Lemma~5.27.  The remaining Lemmas are auxiliary
results on the way to these Lemmas.

These auxiliary lemmas implement some the following principles: in
bounds by $G,h,{\cal A}$ norms, a fluctuation field $\zeta (x)$
contributes $C_{L}$ and a field $\phi$ contributes $O (1)g^{-1/4}$.
In bounds by the $1,{\cal A}$ norms, fluctuation fields and $\phi$
fields contribute $O (1)$.

\vskip.3truecm

\\{\it Lemma 5.1

\\Let $Z$ be a 1-polymer, $Y$ be a $L^{-1}$-polymer  or $\emptyset$,
$Y\subset Z$ and vol$(Z\backslash Y)\ge {1\over 2}$. Choose
$\g=O(1)>0$, $\k=O(1)>0$.  Let $\s$ sufficiently large.  Then for any
$x\in Z$, there exists a constant $O(1)$ depending on $\k,\g,j$ such
that

$$\Vert\phi\Vert_{C^{2}(Z)}^j\le O(1)g^{-{j\over 4}}
e^{\g g\int_{Z\backslash Y}d^3y|\phi(y)|^4}G_\k(Z,\phi)\Eq(5.1)$$
}

\\{\it Proof}

\\This is a simple variant of an analogous lemma in [BDH-est]. Write

$$\phi(x)={1\over {\rm vol}(Z\backslash Y)}\int_{Z\backslash Y}d^3y
(\phi(y) + \phi(x)-\phi(y))
$$

\\and bound

$$\eqalign{|\phi(x)|\le &{1\over {\rm vol}(Z\backslash Y)}\int_{Z\backslash Y}d^3y
|\phi(y)|+{1\over {\rm vol}(Z\backslash Y)}\int_{Z\backslash Y}d^3y
|\phi(x)-\phi(y)|\cr
\le &O(1)\left(\Vert\phi\Vert_{L^4(Z\backslash Y)}+
\Vert\phi\Vert_{Z,1,\s}\right)\cr}
$$

\\where the first term was bounded using the H\"older inequality. The
second term was bounded by writing the difference as the integral of
$\nabla\phi$ and using

$$
|\nabla\phi (x)| \le O (1) \Vert \phi   \Vert_{\D,1,\s},
\qquad 
{\rm for } \ \D \ni x
$$

\\which is the Sobolev embedding theorem, valid for $\sigma >3/2 + 2$.
We also have under the same condition on
$\sigma$
$$\Vert\nabla^{2}\phi\Vert_{C(Z)}\le O(1)\Vert\phi\Vert_{Z,1,\sigma}$$
Hence
$$\eqalign{\Vert\phi\Vert_{C^{2}(Z)}^j\le & O(1)
\left(\Vert\phi\Vert_{L^4(Z\backslash Y)}^j+
\Vert\phi\Vert_{Z,1,\s}^j\right)\cr
\le & O(1)g^{-j/4}e^{\g g\int_{Z\backslash Y}d^3y
|\phi(y)|^4}G_\k(Z,\phi)\cr}$$
where $O(1)$ depends on
$\k,\g,j$. This proves the lemma.

\\For fluctuation fields $\z$, we will have occasion to use the following 
 lemmas

\\Define
$$\tilde G_{\k,\a}(X,\z)=e^{\a\Vert\z\Vert^2_{L^{2}(X)}}G_{\k}(X,\z)
,\quad \a,\k>0\Eq(5.2)$$
$\k$ is O(1) and will be held sufficiently small. The choice of $\a$
is dictated by Lemma 5.3 below.
\vglue.3truecm

\\{\it Lemma 5.2

\\For any $x\in X$
$$|\z(x)|^j\le C_{\a,j}\tilde G_{\k,\a}(X,\z)\Eq(5.3)$$

\\where $$C_{\a,j}=(\a)^{-(j/2)}O(1) $$ 
and $O(1)$ depends  on  $j$ and $\k$. We have isolated out the $\a$
dependance in the bound.
}

\vglue.3truecm

\\{\it Lemma 5.3

$$\int d\m_\G(\z)\tilde G_{\k,\a}(X,\z)\le 2^{|X|}\Eq(5.4)$$
for $\a=\a(L)> 0$ sufficiently small and $\k =O(1)> 0$ sufficiently
small. Here 

$$\a(L)=L^{-(3-2[\phi])}\k' = L^{-(3+\e)/2}\k'$$

\\and $\k' =O(1)> 0$ is held sufficiently small.
}

\\The proof of Lemma 5.2 follows the lines of Lemma~5.1 except that
we replace the $L^4$ norm there by the $L^2$ norm in the appropriate
places and $Y=\emptyset$ and $Z=\Delta \ni x$. The proof of Lemma~5.3
is the same as the one referenced for \equ({2.3}). 

\vglue.3truecm
\\It is convenient, for the control of norms of our polymer activities 
in  intermediate steps, to introduce some new regulators and
some intermediate norms in the following way.

\\Define
$$\hat G_{\k,\a}(X,\z,\phi)=G_\k(X,\z+\phi)G_\k(X,\phi)\tilde G_{\k,\a}(X,\z)
\Eq(5.9)$$
$\hat G_{\k,\a}$ is a regulator. 
\vglue.3truecm

\\{\it Lemma 5.4

$$\int d\m_\G(\z)\hat G_{\k,\a}(X,\z,\phi)\le 2^{|X|}G_{3\k}(X,\phi)
\Eq(5.12)$$
for $\a=\a(L)> 0$ sufficiently small and $\k =O(1)> 0$ sufficiently
small.
}

\vglue.3truecm

\\{\it Proof}: use Cauchy-Schwartz, stability of $G_{\k}$, \equ({2.3})
and Lemma~5.3.

\vglue.3truecm

\\For polymer activity $K(X,\z,\phi)$ define the norms

$$\Vert K(X)\Vert_{h,\hat G_{\k,\a}}=\sup_{\phi,\z}\Vert K(X,\z,\phi)\Vert_h
\hat G_{\k,\a}^{-1}(X,\z,\phi)\Eq(5.10)$$

$$\Vert K(X)\Vert_{h_{*},\tilde G_{\k,\a}}=\sup_{\z}\Vert K(X,\z,0)\Vert_{h_{*}}
\tilde G_{\k,\a}^{-1}(X,\z)\Eq(5.10.1)$$

\\where in \equ(5.10) and \equ(5.10.1) the functional derivatives in
$\Vert K(X,\z,\phi)\Vert_h$ and in $\Vert K(X,\z,0)\Vert_{h_{*}}$ are
computed with respect to the field $\phi$.

\\The norms above are useful because before fluctuation integration
we will encounter activities
$K(X,\z,\phi)$ which are not just functions of $\z+\phi$.

\\The following lemma is a variant of Theorem 1 [BDH-est]
adapted to our purposes.
\vglue.3truecm
\\{\it Lemma 5.5

\\For $V (Y,\phi,\z )= V (Y,\phi+\z ,C,g,\mu)$ or $V (Y,\phi+\z
,C_{L^{-1}},g,\mu)$,

$$\Vert e^{- V(Y,\phi+\z )}\Vert_h\le 2^{|Y|}
e^{-g/4\int_Yd^3x (\phi+\zeta)^4(x)}\Eq(5.5)$$

$$
\Vert e^{-V(Y,\phi+\z)}\Vert_{h_{*}}\le  2^{|Y|}\Eq(5.lem5.1)
$$

\\for $\e >0$ sufficiently small, and, for the second bound, depending
on $L$.
}

\vglue.3truecm
\\{\it Proof}:
$${\tilde V}(\D,\phi)=V(\D,\phi,C_{L^{-1}},g,\m)=g\int_\D 
d^3x:\phi^4:_{C_{L^{-1}}}(x)+
\m\int_\D d^3x:\phi^2:_{C_{L^{-1}}}(x)$$
 Undo the Wick ordering. Wick constants 
are finite and $O(1)$. Recall  from the initial Hypothesis that 
$g=O(\e)$ and $\vert \m\vert \le O(\e^2)$,
and $h=c\e^{-1/4}$, with $c$ small enough. Hence

$${\tilde V}(\D,\phi))=g\int_\D 
d^3x\ \phi^4(x)-O(1)g
\int_\D d^3x\ \phi^2(x) -O(\e)$$

$${\tilde V}(\D,\phi)-{g\over 2}\int_\D 
d^3x\ \phi^4(x)={g\over 2}\int_\D 
d^3x\ \phi^4(x)-O(1)g
\int_\D d^3x\ \phi^2(x) -O(\e)=$$
$$={g\over 2}\int_\D 
d^3x\ \left((\phi^2(x)-O(1))^2-O(1)\right)\ge -O(\e)$$
Hence
$$e^{-\tilde V(\D,\phi)}
\le (1+O(\e))e^{-g/2\int_\D d^3x\phi^4(x)}
$$
Compute now the derivatives $D^k\ e^{-\tilde V}$.
We get for $1\le k\le n_0$
$${h^k\over k!}\left\Vert(D^k\ e^{-\tilde V})(\D,\phi)\right\Vert
\le c^{k}\sum_{j=1}^k {1\over j!}\sum_{1\le l_i \le 4 \atop\sum l_i = k}
\prod_{i=1}^j {(\e ^{-1/4})^{l_i}\over l_i !}
\left\Vert D^{l_i}\tilde {V}(\D,\phi)
\right\Vert e^{-\tilde {V}(\D,\phi)}$$
$$ \le   c^{k}\sum_{j=1}^k {1\over j!}\Bigl(\sum_{1\le l\le 4}{(\e ^{-1/4})^{l}
\over  l!}\left\Vert D^{l}\tilde {V}(\D,\phi)\right\Vert\Bigr)^{j}
 e^{-\tilde {V}(\D,\phi)}$$
$$\le
  O(1)c^{k} e^{-{g\over 2}\int_{\D}d^{3}x\phi^{4}(x)}
e^{\sum_{1\le l\le 4}{(\e ^{1/4})^{l}
\over  l!}\left\Vert D^{l}\tilde {V}(\D,\phi)\right\Vert}$$
Take the expression for $\tilde V$ where the Wick ordering has been undone.
Then it is easy to see that
$$\sum_{1\le l\le 4}{(\e ^{-1/4})^{l}
\over  l!}\left\Vert D^{l}\tilde {V}(\D,\phi)\right\Vert
\le   O(1)\int_{\D}d^{3}x\sum_{j=0}^3 \left\vert\e ^{1/4} \phi
\right\vert ^j \le {g\over 4}\int_{\D}d^{3}x\phi^{4}(x) + O(1) $$
Hence 

$${h^k\over k!}\left\Vert(D^k\ e^{-\tilde V})(\D,\phi)\right\Vert
\le O(1)c^{k} e^{-{g\over 4}\int_{\D}d^{3}x\phi^{4}(x)} $$

\\The sum over $k$ is $O(c)$ if $c$ is small enough.  The proof of
\equ({5.5})follows easily. The proof of \equ({5.lem5.1}) follows
the same lines but we must take $\e$ sufficiently small depending on
$L$.

\vglue.3truecm

\\{\it Lemma 5.6

\\Let $p_{g}(\D,\z,\phi), p_{\mu} (\D, \z, \phi)$ be as given in
\equ(4.8b).  $h=c\e^{-1/4}$, $g,\mu $ as in the inductive
hypothesis, and $h_{*}$ as defined earlier.
Then for any $\a>0$, $\k=O(1)>0$, $\xi=O(1)>0$, $0\le s <
1$

\\$$\Vert p_{g}(\D,\z,\phi)\Vert_h\le C_\a \e^{1/4}(1-s)^{-3/4}
\tilde G_{\k,\a}(\D,\z)G_{\k}(\D,\phi)e^{g(1-s)\xi\int_\D
d^3x\ \phi^4(x)}\Eq(5.6)$$

\\$$\Vert p_{\mu}(\D,\z,\phi)\Vert_h\le C_\a \epsilon^{7/4-\delta
}(1-s)^{-1/2} \tilde G_{\k,\a}(\D,\z)G_{\k}(\D,\phi)e^{g(1-s)\xi\int_\D
d^3x\ \phi^4(x)}\Eq(5.6b)$$

$$\Vert p_{g}(\D,\z,0)\Vert_{h_{*}}\le C_{\a,L} \e \tilde G_{\k,\a}(\D,\z)
\Eq(5.6.1)
$$

$$\Vert p_{\mu}(\D,\z,0)\Vert_{h_{*}}\le C_{\a,L} \e^{2-\delta }  \tilde G_{\k,\a}(\D,\z)
\Eq(5.6.2)
$$
}

\vglue.3truecm
\\{\it Proof}

\\Undoing the Wick ordering we have

$$p_{g}(\D,\z,\phi)=g
\int_\D d^3x
\sum_{j=1}^3a_j\z^{4-j}(x)\phi^j(x)$$

\\where the constants $a_j$ are $O(1)$.

$$h^k\left\vert D^kp_{g}(\D,\z,\phi;f^{\times k})\right\vert\le
\e^{-k/4}g\int_\D d^3x\ 
\sum_{j=1}^3a_j|\z^{4-j}(x)||\phi^{j-k}(x)|\Vert f\Vert^{\times k}_{C^2(\D)}$$

\\Note that since $p_{g}$ is a third degree polynomial in $\phi$,
derivatives with $k>3$ vanish. Moreover on the right hand side $j\ge
k$.  Now use Lemma~5.1, 5.2 to get

$$h^k\left\Vert D^kp_{g}(\D,\z,\phi)\right\Vert\le C_\a g\e^{-k/4}
g^{-{3-k\over 4}}(1-s)^{-3/4}
\tilde G_{\k,\a}(\D,\z)G_{\k}(\D,\phi)e^{g(1-s)\xi\int_\D
d^3x\ \phi^4(x)}$$

\\Use $g=O(\e)$, multiply by ${1\over k!}$ and take sum over $k$  to
obtain \equ(5.6). The remaining parts are obtained along the same
lines.

\vglue.3truecm
\\Define $p (s) = p (s,\Delta ,\phi,\zeta )$ by 

$$
p (s) = sp_{g}+s^{2}p_{\mu} \Eq(5.7a)
$$

\\Then $r_{1} = r_{1} (\Delta ,\phi ,\z)$ defined by \equ({4.9}) is
given by

$$r_{1} =
{1\over 2}\int_0^1ds(1-s)^{2}e^{-p(s)-\tilde{V}}
\big(-p' (s)^{3}+6p' (s)p_{\mu }\big)\Eq(5.7)$$

\\with $p'(s)={d\over ds}p (s)=p_{g}+2sp_{\mu }$ and $p'' (s)=2p_{\mu}$.

\vglue.3truecm

\\{\it Lemma 5.7

$$\Vert r_{1}(\D)\Vert_{h,\hat G_{\k,\a}} \le C_{\a}\e^{3/4}\Eq(5.8)$$

$$\Vert r_{1}(\D)\Vert_{h_{*},\tilde G_{\k,\a}}\le C_{\a,L}\e^{3-\delta }\Eq(5.8.1)$$
}

\vglue.3truecm
\\{\it Proof}

\\The hypotheses for $g,\mu$ also hold for $sg,s^{2}\mu$ and for
$[1-s]g,[1-s^{2}]\mu$. Since $V+p (s) = V_{1} (s)+V_{2} (s)$ with

$$
V_{1} (s) = V (\Delta ,\phi +\z, C, sg, s^{2}\mu),
\qquad 
V_{2} (s) = V (\Delta ,\phi, C_{L^{-1}}, [1-s]g, [1-s^{2}]\mu)
$$
 
$$\Vert r_{1}(\D,\z,\phi)\Vert_h\le
{1\over 2}\int_0^1ds(1-s)^{2}\Vert e^{-V_{1}(s)}\Vert_h
\Vert e^{-V_{2} (s)}\Vert_h
\bigg(
\Vert p' (s)\Vert_{h}^{3} +
6\Vert p' (s)\Vert_{h} \Vert p_{\mu }\Vert_h
\bigg)
$$

\\By Lemma~5.5, the exponential terms are bounded by $4\exp (-
(g[1-s]/4)\int \phi^{4} )$.  Use Lemma~5.6 choosing for the
regulators the constants $\k\over 4$ and $\a\over 3$ to obtain

$$\Vert r_{1}(\D,\z,\phi)\Vert_h\le
C_{\a}\e^{3/4}\hat G_{\k,\a}\int_0^1ds(1-s)^{2}(1-s)^{-{9\over 4}}
e^{-{g\over 4}(1-s)\int_\D
d^3x\ \phi^4(x)}e^{g(1-s)3\xi\int_\D
d^3x\ \phi^4(x)}$$

\\Choose $0<\xi<{1\over 12}$ to get \equ(5.8). Equation \equ(5.8.1)
is proved similarly.

\vglue.3truecm

\\{\it Lemma 5.8

\\Consider $P (\lambda )$ given in \equ(4.9). Then for $C$
independent of $\e$, but dependent on $L$, 

$$\Vert P\Vert_{h,\hat G_{\k,\a},\AA}\le 
C_{L}\vert\lambda\e\vert^{1/4}
\qquad {\rm for } \ |\lambda \e^{1/4}| \le 1
\Eq(5.13)
$$

$$\Vert P\Vert_{h_{*},\tilde G_{\k,\a},\AA}
\le C_{L}\vert\lambda\e^{1-\delta /2}\vert   
\qquad {\rm for } \ |\lambda \e^{1-\delta /2}| \le 1
\Eq(5.13.1)$$
}

\vglue.3truecm
\\{\it Proof}

\\\equ(5.13) and \equ(5.13.1) are immediate from Lemma~5.7, noting
that $\lambda\e^{1-\delta /2}$ in \equ({5.13.1}) is the largest of
the several combinations of $\lambda ,\e$ that arise.  It comes from
the term involving $\mu\lambda^{2}$.

\vglue.5truecm 

\\{\it Estimates for $Qe^{-V}$}

\\We now turn to the estimate of $Qe^{-V}$. From \equ(4.19.1)
$$Q(C,{\bf w},g)=g^2\sum_{m=1}^3 a_mQ^{(m,m)}(C,w^{(4-m)})\Eq(5.18)$$
where the $a_m$ are numerical coefficients and the $Q^{(m,m)}$ are given
in \equ(4.18). Under an iteration, see Proposition~4.1, we have
$$w^{(p)}\rightarrow w^{(p)'}=v^{(p)}+w_L^{(p)}$$ where $p=1,2,3$ and
the $v^{(p)}$ are given in Proposition~4.1.  Starting with
$w_0^{(p)}=0$ we get after $n$ iterations
$$w_n^{(p)}=\sum_{j=0}^{n-1}v_{L^j}^{(p)}\Eq(5.19)$$
We need to first estimate $w_n^{(p)}$ and the limit $\lim_{n\rightarrow\io}
w_n^{(p)}$ under appropriate norms.

\vglue.3truecm

\\We consider Banach spaces ${\cal W}_p$ of measurable functions with
norms $\Vert\cdot\Vert_p$, $p=1,2,3$, defined as follows
$$\Vert f\Vert_p= ess.\sup_x\left(|x|^{6p+1\over
4}|f(x)|\right)\Eq(5.20)$$ We define the Banach space ${\cal
W}_{1}\times {\cal W}_{2}\times{\cal W}_{3}$ consisting of vectors
$\bf w$ with the norm

$$\Vert{\bf w}\Vert = \sup_{p} \Vert w_{p}\Vert_{p}$$

\\Then we have

\vglue.3truecm

\\{\it Lemma 5.9

\\For $L$ sufficiently large and $\e>0$ sufficiently small there
exists a constant $c=O (1)$ such that

$$\Vert {\bf w}_n\Vert\le c/4\ \ \forall n$$

\\and

$$\Vert{\bf w}_{n+1}-{\bf w}_{n}\Vert\le c/8\ \  L^{-n/4} $$

\\so that ${\bf w}_n\rightarrow {\bf w}_*$ in the norm $\Vert\cdot\Vert$,
and
 
$$\Vert {\bf w}_*\Vert\le c/4$$
}

\vglue.3truecm
\\{\it Proof}

\\Let us note first some weak uniform (in $L$) bounds. Recall that
$[\phi]= {3-\e\over 4}$.

$$|\G_{L}(x)|\le O(1)|x|^{-2[\phi]}\Eq(5.21)$$

$$|C(x)|\le O(1)\Eq(5.22)$$

\\To see this observe that from the definition of $\G$ (see Section 1)

$$|\G_{L}(x)| \le \int_{0}^{\infty}
{dl\over l} l^{-2[\phi]}\left|u\left({x\over
{l}}\right)\right|$$

\\Let $x\ne 0$. Then, using support properties of $u$,

$$|\G_{L}(x)|\le \int_{|x|}^{\infty}
{dl\over l} l^{-2[\phi]}\ \Vert u\Vert_\io = {2\over 3-\e}\ 
{1\over |x|^{2[\phi]}}\ \Vert u\Vert_\io$$

\\which proves \equ(5.21). To prove \equ(5.22) recall

\\$$|C(x)|\le
\int_{1}^{\io}{dl\over l} l^{-2[\phi]}\left|u\left({x\over
{l}}\right)\right|
\le O (1) \Vert u\Vert_\io
$$

\\which proves \equ(5.22).

\vskip.3truecm

\\For $p=1,2,3$,

$$\eqalign{
\vert v^{(p)}(x) \vert &= \vert C_{L}^p(x)-C^p(x) \vert\cr
&\le p\sup_{1\le q \le p}\vert  \G_{L}(x) \vert^{q} 
\vert C (x)\vert^{p-q}
\le O (1)\cases{
\vert x \vert^{-p2[\phi]} & \cr
0 & if $\vert x\vert \ge 1$
}
}
$$

\\where we exploited $C_{L}=\G_{L}+C$ and $\G_{L} (x) = 0$ if $\vert
x\vert \ge 1$ and \equ({5.21}) and \equ({5.22}). Hence 

$$
\Vert v^{(p)} \Vert_{p}\le O (1)
$$

\\and by scaling $x = L^{j}x'$

$$\Vert v_{L^j}^{(p)}\Vert_p \le
L^{jp2[\phi]}L^{-j{(6p+1)\over 4}}\Vert v^{(p)}\Vert_p
\le\ c_{p}/8\ \ L^{-j/4} \Eq(5.24)$$

\\Define the constant $c=\sup_{p} c_{p}$.  Then the above estimates
lead immediately to the proof of Lemma~5.9, because of \equ({5.19}).

\vglue.3truecm

\\{\it Lemma 5.10

\\$Q(X,\phi)e^{- V(X,\phi)}$satisfies the bounds
$$\left\Vert Qe^{- V}\right\Vert_{h,G_\k,\AA_{p}}
\le C_{p}\e^{1/2}\Eq(5.25)$$
$$\left\vert Qe^{- V}\right\vert_{h_{*},\AA_{p}}
\le C_{p}\e^{2}\Eq(5.25.1)$$
}

\vglue.3truecm
\\{\it Proof}
$$Qe^{- V}=g^2\sum_{m=1}^3a_mQ^{(m,m)}(C,w^{(4-m)})e^{- V}$$
$$\left\Vert Q(X,\phi)e^{-V(X,\phi)}\right\Vert_h\le
g^2\sum_{m=1}^3|a_m|\left\Vert Q^{(m,m)}(C,w^{(4-m)},X,\phi)\right\Vert_h
\left\Vert e^{-V(X,\phi)}\right\Vert_h\Eq(5.26)$$
Here $X$ is a small set, because of the support property of $Q$.
The last factor in the sum will be estimated by Lemma~5.5.
>From \equ(4.18)
$$Q^{(3,3)}(\tilde X,\phi;C,w^{(1)})={1\over 2}
\int_{\tilde X}d^3xd^3y:\phi^3(x)\phi^3(y):_C
w^{(1)}(x-y)$$
Undo the Wick ordering, which produces lower order terms with
finite coefficients.

\\Apply $h^kD^k$ with $h=c\e^{-1/4}$ and use Lemmas 5.1, 5.9
and $g=O(\e)$. We get
$${h^k\over k!}\left\Vert D^kQ^{(3,3)}(\tilde X,\phi;C,w^{(1)})\right\Vert\le
O(1) g^{-3/2}\Vert w^{(1)}\Vert_1
\int_{\tilde X}d^3xd^3y{1\over |x-y|^{7/4}}
G_{\k/4}(X,\phi)e^{g/4\int_Xd^3x\phi^4(x)}\Eq(5.27)$$
Next turn to $Q^{(m,m)}$, $m=1,2$, again in \equ(4.18)
$$Q^{(m,m)}(\tilde X,\phi;C,w^{(4-m)})=-{m^2\over 2}\sum_{\m,\n=1}^3
\int_0^1ds_1ds_2
\int_{\tilde X}d^3xd^3y$$
$$(x-y)_\m(x-y)_\n w^{(4-m)}(x-y)
:(\phi^{m-1}\nabla_\m\phi)(y+s_1(x-y))
(\phi^{m-1}\nabla_\n\phi)(y+s_2(x-y)):_C$$
We consider in turn the cases $m=2,1$. We 
apply $h^kD^k$ with $h=a\e^{-1/4}$ and use Lemmas 5.1-5.9
and the Sobolev inequality to dominate the $\nabla\phi$
pointwise by the large field regulators.
$${h^k\over k!}\left\Vert D^kQ^{(2,2)}(\tilde X,\phi;C,w^{(2)})\right\Vert\le$$
$$\le O(1) g^{-1/2}\Vert w^{(2)}\Vert_2\sum_{\m,\n=1}^3
\int_{\tilde X}d^3xd^3y{|(x-y)_\m||(x-y)_\n |\over|x-y|^{13/4}}
G_{\k/4}(X,\phi)e^{g/4\int_Xd^3x\phi^4(x)}$$
$$\le cg^{-1/2}G_{\k/4}(X,\phi)e^{g/4\int_Xd^3x\phi^4(x)}\Eq(5.28)$$
We can estimate in the same way the case $m=1$. Note that
$$\int_{\tilde X}d^3xd^3y{|(x-y)_\m||(x-y)_\n |\over|x-y|^{19/4}}
\le O(1)$$
We have
$${h^k\over k!}\left\Vert D^kQ^{(1,1)}(\tilde X,\phi;C,w^{(3)})\right\Vert\le
cG_{\k/4}(X,\phi)e^{g/4\int_Xd^3x\phi^4(x)}\Eq(5.29)$$
>From \equ(5.27), \equ(5.28) and \equ(5.29) we get
$$g^2\left\Vert Q^{(m,m)}(C,w^{(4-m)})\right\Vert_h
\le c\e^{1/2}G_{\k/4}(X,\phi)e^{g/4\int_Xd^3x\phi^4(x)}\Eq(5.30)$$
We estimate the r.h.s.of \equ(5.26) using \equ(5.30) and Lemma~5.5.
$$\left\Vert Q(X)e^{-V(X)}\right\Vert_{h,G_\k}\le O(1)\e^{1/2}
\Eq(5.31)$$
Since the $Q$ are supported on small sets, we get from \equ(5.31)
$$\left\Vert Qe^{-V}\right\Vert_{h,G_\k,\AA_{p}}\le C_{p}\e^{1/2}
\Eq(5.32)$$
This proves \equ(5.25). 
To prove \equ(5.25.1) we estimate
the r.h.s. of \equ(5.26) at $\phi=0$ after undoing the Wick ordering,
set $h=h_{*}$, using Lemma~5.1.
Lemma~5.10 has been proved.
\vglue.3truecm
\\It is useful to state the following lemma for $Qe^{-V}$
treated as a function of $\z, \phi$
\vglue.3truecm

\\{\it Lemma 5.11

$$\left\Vert Qe^{-V}\right\Vert_{h,\hat G_{\k,\a},\AA_{p}}\le
C_{p}\e^{1/2} \Eq(5.33)$$

$$\left\Vert Qe^{-V}\right\Vert_{h_{*},\tilde G_{\k,\a},\AA_{p}}\le
C_{p}\e^{2} \Eq(5.34)$$
}

\vglue.3truecm
\\{\it Proof}

\\\equ(5.33) follows from \equ(5.25), since $\hat G_{\k,\a}\ge G_\k$.
For \equ(5.34) we bound derivatives with respect to $\phi$
of $(Qe^{-V})(\z+\phi, X)$ at $\phi=0$
$$\left\Vert Q(X,\z)e^{-V(X,\z)}\right\Vert_{h_{*}}\le
g^2O(1)
\sum_{m=1}^3\left\Vert Q^{(m,m)}(C,w^{(4-m)},X,\z)\right\Vert_{h_{*}}
\tilde G_{\a/2}(X,\z)$$
since 
$$\left\Vert e^{-V(X,\z)}\right\Vert_{h_{*}}\le  2^{|X|} $$
and $X$ is a small set. Now we proceed as in Lemma~5.10, but dominate
the $\z$ using Lemma~5.2, to complete the proof.

\vglue.3truecm

\\We now control the perturbative flow coefficients $a,b$ given in
\equ(4.39).

\vglue.3truecm

\\{\it Lemma 5.12
$$a=O(\log L),
\qquad b=O(L^{3/2}),
\qquad \int d^3x\,v^{(4)}(x) = O (L^{3})
\Eq(5.35)$$

\\for $L\rightarrow \infty$ and $\epsilon =o (L)$.
}

\vglue.3truecm
\\{\it Proof}

\\From \equ(4.54), for $p=2,3$, using $C_L=\G_L+C$

$$v^{(p)}=C^p_L-C^p=\G_L(\G_L^{p-1}+p\G_L^{p-2}C+\d_{p,3}3C^2)\Eq(5.36)$$

\\with pointwise multiplication.  The common factor of $\G_{L}$
implies that $v^{(p)} (x)$ has support in the unit ball. This,
together with \equ(5.22), implies that the leading divergence is in
the contribution from $\G_{L}^{p}$. Therefore is sufficient to
calculate

$$
I^{(p)} = \int d^3x\,(\G_L(x))^{p}
$$

\\Recall that:

$$\G_{L}(x) = \int_{L^{-1}}^{1}
{dl\over l} l^{-2[\phi]}\ u\left({x\over l}\right) \Eq(5.37)
$$

\\Let $p=2$ so that

$$I^{(p)} = 2! \int_{L^{-1}\le l_{1}\le l_{2}\le 1} {dl_1\over l_1}
{dl_2\over l_2} (l_1l_{2})^{-2[\phi]} \int d^3x\, u\left({x\over
l_1}\right)u\left({x\over l_2}\right)$$

\\Suppose that $L=2^{N-1}$ for some integer $N$ and break up the range
of integration into disjoint regions

$$
R_{n} = \{2^{-n-1}< l_{1}\le l_{2} \le 2^{-n} \} \qquad n= 0,....,N-1
$$

\\so that

$$
I^{(p)} = \sum_{n=0}^{N-1} I^{(p)} (R_{n})
$$

\\where

$$I^{(p)}(R) = 2! \int_{R} {dl_1\over l_1} {dl_2\over l_2}
(l_1l_{2})^{-2[\phi]} \int d^3x\, u\left({x\over
l_1}\right)u\left({x\over l_2}\right)$$

\\By scaling $l_{1},l_{2},x$ by $2^{n}$, $I^{(p)}
(R_{n})=2^{2p[\phi]-3}I (R_{0})$.  Therefore

$$
I^{(p)} = I^{(p)} (R_{0})\sum_{n=0}^{N-1} 2^{(2p[\phi]-3)n}
= O (1)\sum_{n=0}^{N-1} 2^{(2p[\phi]-3)n}
= O (1) {1-L^{2p[\phi]-3}\over 1-2^{2p[\phi]-3}}
$$

\\This is also true for $p=3,4$, by the same argument with the
appropriate multiple integral expression for $I^{(p)} (R_{0})$. For
$p=2$, $2p[\phi]-3 = -4\e \rightarrow 0$ and therefore $I^{(p)} = O
(\log L)$ by L'Hospital's rule.  For $p=3$, $2p[\phi]-3 \rightarrow=
3/2$ and therefore $I^{(p)} = O (L^{3/2})$, etc.   \equ({5.35}) is proved.

\vglue.3truecm

\\{\it Lemma 5.13

$$\left\Vert Q(e^{-V}-e^{-\tilde V})\right\Vert_{h,\hat G_{\k,\a},\AA_{p}}
\le C_{p}\e^{3/4}
\Eq(5.43)$$

$$\left\Vert Q(e^{-V}-e^{-\tilde V})\right\Vert_{h_{*} ,{\tilde G}_{\k,\a}
,\AA_{p}} \le C_{p}\e^{3} \Eq(5.44)$$
}

\vglue.3truecm
\\{\it Proof}

\\$Q(X)$ is supported on connected polymers with size $|X|\le 2$.
Without loss of generality we do the estimates for $|X|=1$.
We can write:
$$Q(\D,\z+\phi)(e^{-V(\D,\z+\phi)}-e^{-\tilde V(\D,\phi)})
=Q(\D,\z+\phi)e^{-{1\over 2}V(\D,\z+\phi)}
\int_0^{1/2}ds\, p(\D,\z,\phi)
e^{-({1\over 2}-s)V(\D,\z+\phi)-s\tilde V(\D,\phi)}
+$$
$$+\int_{1/2}^1ds\,Q(\D,\z+\phi) p(\D,\z,\phi)
e^{-(1-s)V(\D,\z+\phi)-s\tilde V(\D,\phi)}$$
whence
$$\left\Vert Q(\D,\z+\phi)(e^{-V(\D,\z+\phi)}-e^{-\tilde V(\D,\phi)})
\right\Vert_h\le O(1)(A+B)\Eq(5.45.1)$$
where
$$A=\left\Vert Q(\D,\z+\phi)e^{-{1\over 2}V(\D,\z+\phi)}
\right\Vert_h 
\int_0^{1/2}ds\, \Vert p(\D,\z,\phi)\Vert_h\Vert
e^{-s\tilde V(\D,\phi)}\Vert_h$$
and
$$B=\int_{1/2}^1ds\,\Vert Q(\D,\z+\phi)\Vert_h\Vert p(\D,\z,\phi)\Vert_h
\Vert
e^{-s\tilde V(\D,\phi)}\Vert_h$$
To estimate $A$, use Lemma~5.11, still true for $V$ replaced by ${1\over2}V$,
Lemma~5.6, \equ(5.6) with $1-s$ replaced by $s$, $\d={1\over8}$, and lemma
5.5 with $g$ replaced by $sg$. Observe that $s^{-3/4}$ is integrable. We get
$$A\le C_\a\e^{3/4}\hat G_{\k,\a}(\D,\z,\phi)\Eq(5.46.1)$$
To estimate B we use again Lemma~5.5 with $g$ replaced by $sg$. We 
estimate $\Vert p\Vert_h$ using Lemma~5.6, \equ(5.6)
with $1-s$ replaced by $s$, $\d={1\over16}$. We estimate $\Vert Q
(\D,\z+\phi)\Vert_h$ as in the proof of Lemma~5.10 with the following 
change. Expand out polynomials in $\z+\phi$, and dominate the $\z$ using
Lemma~5.3. We then get a modified estimate \equ(5.30) replacing $g$ with
${s\over 4}g$, and $G_\k(X,\phi)$ with $\hat G_{\k,\a}(X,\z,\phi)$
together with an overall multiplicative factor $C_\a s^{-3/2}$
which is integrable in the range under consideration. 
Then we use lemma
5.5 with $g$ replaced by $sg$. Putting all this together we get
$$B\le C_\a\e^{3/4}\hat G_{\k,\a}(\D,\z,\phi)\Eq(5.47.1)$$
Using \equ(5.46.1) and \equ(5.47.1) we get for \equ(5.45.1)
$$\left\Vert Q(\D,\z+\phi)(e^{-V(\D,\z+\phi)}-e^{-\tilde V(\D,\phi)})
\right\Vert_h\le C_\a\e^{3/4}\hat G_{\k,\a}(\D,\z,\phi)\Eq(5.48.1)$$
It is easy to show that the same estimate holds if $|X|=2$, connected.
Hence
$$\left\Vert Q(e^{-V}-e^{-\tilde V})\right\Vert_{h,\hat G_{\k,\a},\AA}
\le C_{\a,L}\e^{3/4}$$
This proves \equ(5.43). The proof of \equ(5.44) is similar except that we 
have only fluctuation fields $\z$ to dominate using Lemma~5.3.
This proves Lemma~5.13.

\vglue.3truecm

\\{\it Lemma 5.14

\\$K (\lambda )$ given by \equ(4.10), satisfies the bounds

$$\Vert K (\lambda )\Vert_{h,\hat G_{\k,\a},\AA}
\le C_\a \vert \lambda \e^{1/4-\eta /3}\vert^{2}
\qquad {\rm for } \ \vert \lambda \e^{1/4-\eta /3}\vert \le 1
\Eq(5.55)$$

$$\Vert K (\lambda )\Vert_{h_{*},\tilde G_{\k,\a} ,\AA}\le 
C_{\a,L}\vert \lambda \e^{11/12-\eta /3}\vert^{2}
\qquad {\rm for } \ \vert \lambda \e^{11/12-\eta /3}\vert \le 1
\Eq(5.56)$$
}

\vglue.3truecm
\\{\it Proof}

\\This follows from lemmas~5.11 and 5.13 and the hypotheses
\equ({55.2}) and \equ({55.3}) on $R$.  The $\lambda \e^{1/4-\eta
/3}$ and $\lambda \e^{11/12-\eta /3}$ originate from $\lambda^{3}R$
contributions.

\vglue.3truecm

\\{\it Lemma 5.15

\\For any polymer activity $K$:

$$\Vert K(X,\z)\Vert_{h_{*}}\le O(1) \tilde G_{\a,\k}(X,\z)\left[
\vert K(X)\vert_{h_{*}}+h^{-n_0}h_{*}^{n_{0}}\Vert
K(X)\Vert_{h,G_\k}\right]\Eq(5.45)$$

$$\Vert K(Y,\phi)\Vert_{h}\le O(1) 
e^{\g g\int_{Z\backslash Y}d^3y|\phi(y)|^4}G_\k(Z,\phi)
\left[
\vert K(Y)\vert_{h}+L^{-n_0[\phi]}\Vert
K(Y)\Vert_{L^{[\phi]}h,G_\k}\right]\Eq(5.45b)$$

$$\vert K^\sharp(X)\vert_{h_{*}}\le O(1) 2^{|X|}\left[ \vert
K(X)\vert_{h_{*}}+h^{-n_0}h_{*}^{n_{0}}\Vert
K(X)\Vert_{h,G_\k}\right]\Eq(5.46)$$

\\where $\tilde G_{\a,\k}$ is as defined in \equ(5.2), and $n_0$ is
the maximum number of derivatives appearing in the definition of
Kernel and $h$ norms. In \equ({5.45b}), $Y,Z,\gamma$ are as described
in Lemma~5.1.

} \vglue.3truecm

\\Recall that $n_{0}=9$. The superscript $\sharp$ stands for
$d\m_{\G}(\z)$ integration. $\a$ is chosen as in Lemma 5.3. Note that
we have then

$$
\a =\a(L)= {\k'\over h_{*}^2} \Eq(5.lem5.16-1)
$$

\vglue.3truecm
\\{\it Proof}

\\First observe:
$${1\over n_0!}\Vert(D^{n_0}K)(X,\z)\Vert\le
h^{-n_0}\tilde G_{\a,\k}(X,\z)\Vert K(X)\Vert_{h,G_\k}\Eq(5.46.2)$$
since $G_{\k}(X,\z)\le\tilde G_{\a,\k}(X,\z)$.

\\Now let $n<n_0$. We expand in Taylor series with remainder
$$(D^{n}K)(X,\z;f^{\times n})=
\sum_{m=0}^{n_0-n-1}{1\over m!}
(D^{n+m}K)(X,0;f^{\times n},\z^{\times m})+$$
$$+{1\over (n_0-n-1)!}\int_0^1ds(1-s)^{n_0-n-1}
(D^{n_0}K)(X,s\z;f^{\times n},\z^{\times n_0-n})
$$
whence
$$\Vert (D^{n}K)(X,\z)\Vert \le
\sum_{m=0}^{n_0-n-1}{1\over m!}\Vert\z\Vert^{m}_{C^2(X)}
\Vert (D^{n+m}K)(X,0)\Vert+$$
$$+{1\over (n_0-n-1)!}\int_0^1ds(1-s)^{n_0-n-1}
\Vert\z\Vert^{n_0-n}_{C^2(X)}G_{\k}(X,s\z)
\Vert (D^{n_0}K)(X)\Vert_{G_{\k}}
$$
By Lemma~5.2, with $\z$ replaced by $\sqrt{1-s^{2}}\z$, and 
\equ({5.lem5.16-1}),

$$h_{*}^{p}\Vert\z\Vert^p_{C^2(X)}\le O(1) 
\tilde G_{\a,\k}(X,\sqrt{1-s^2}\,\z)
{1\over(1-s^2)^{p/2}}$$

\\where O(1) depends on $\k'$ and $p$ and $0\le s<1$.  With $s=0$ this
bound is applied to the terms in the sum over $m$.  For the Taylor
remainder term we note that $(1-s)^{(n_0-n)/2-1}$ is integrable since
$n_0>n$.  Hence:

$${h_{*}^{n}\over n!}\Vert(D^{n}K)(X,\z)\Vert\le
O(1) \tilde G_{\a,\k}(X,\z)\left[
\left(\sum_{m=0}^{n_0-n-1}{(n+m)!\over m!n!}\right)
\vert K(X)\vert_{h_{*}}+\right.$$
$$+\left.
{n_0!h^{-n_0}h_{*}^{n_{0}} \over n!(n_0-n-1)!}
\Vert K(X)\Vert_{h,G_\k}\right]\le$$
$$\le O(1) \tilde G_{\a,\k}(X,\z)
\left[
\vert K(X)\vert_{h_{*}}+h^{-n_0}h_{*}^{n_{0}}\Vert K(X)\Vert_{h,G_\k}\right]
\Eq(5.46.3)$$

\\Summing \equ(5.46.3) over $0\le n\le n_0-1$ and adding
\equ(5.46.2) after multiplication by $h_{*}^{n_{0}}$ proves
\equ(5.45).  Equation \equ(5.46) follows from \equ(5.45) using
Lemma~5.3. Equation \equ(5.45b) is proved in the same way as
\equ(5.45) with Lemma~5.1 in the place of Lemma~5.2, $h_{*},h$
replaced by $h,Lh$.  This proves Lemma~5.15.

\vglue.3truecm

\\The next lemma gives bounds on $\SS (\lambda ,K)=\RR\circ {\cal
B} (\lambda,K)$ given in \equ(4.13).

\vglue.3truecm

\\{\it Lemma 5.16

\\For any $q>0$, there exists $c_{L}$ such that, for $L$ large,

$$\Vert\SS(\lambda,K)^\natural\Vert_{h,G_{\k},\AA_{p}}\le 
q
\qquad {\rm when } \   |\lambda \e^{1/4-\eta /3}|\le c_{L}
\Eq(5.57) $$

$$\vert\SS(\lambda,K)^\natural\vert_{h_{*},\AA_{p}}\le q
\qquad {\rm when } \   |\lambda \e^{11/12-\eta/3}| \le c_{L}
\Eq(5.58)$$

\\When $R=0$ we may set $\eta =0$ in \equ({5.57}) and replace
$\lambda \e^{11/12-\eta/3}$ by $\lambda \e^{1-\d/2}$ in \equ({5.58}).
}

\vglue.3truecm
\\{\it Proof}

\\From the definition of reblocking (see \equ(4.13)) and subsequent 
rescaling

$$
\eqalign{
\Vert\SS (\lambda ,K) (Z,\phi,\z) \Vert_{h} \le  
&\sum_{N+M\ge 1}{1\over N !M!}
{\sum}_{(X_{j}),(\Delta_{i})\rightarrow LZ }
\Vert e^{-\tilde{V}(X_{0},\phi_{L^{-1}})} \Vert_{h}
\cr
&{\prod}_{j=1}^{N} 
\Vert K(\lambda, X_{j},\phi_{L^{-1}},\z_{L^{-1}}) \Vert_{h}
{\prod}_{i=1}^{M}\Vert 
P (\lambda ,\Delta_{i},\phi_{L^{-1}},\z_{L^{-1}})
\Vert_{h}
}
\Eq(5.19-1)$$

\\We rewrite $\tilde{V}(X_{0},\phi_{L^{-1}})=
\tilde{V}_{L}(L^{-1}X_{0},\phi)$ and apply Lemma~5.5 (the rewriting
gives a better bound by saving factors of $2$),

$$\eqalign{\Vert\SS (\lambda ,K) (Z,\phi,\z) \Vert_{h} 
\le
&2^{|Z|} \hat G_{\k,\a}(LZ,\z_{L^{-1}},\phi_{L^{-1}})
\sum_{N+M\ge 1}{1\over N !M!}
{\sum}_{(X_{j}),(\Delta_{i})\rightarrow LZ }
\cr
&
{\prod}_{j=1}^{N}\Vert 
K(\lambda, X_{j}) 
\Vert_{h_L,\hat G_{\k,\a}}
{\prod}_{i=1}^{M}
\Vert   \
P (\lambda ,\Delta_{i})
\Vert_{h_L,\hat G_{\k,\a}}
\cr
}
\Eq(5.19-2)$$

\\where

$$h_L=L^{-(3-\e)/4}h$$

\\Using Lemma~5.4 and $G_{3\k}(LZ,\phi_{L^{-1}}) \le
G_{\k}(Z,\phi)$ for $L$ large,

$$\eqalign{
\Vert\SS(\lambda,K)^\natural(Z)\Vert_{h,G_{\k}} \le
&
2^{2|Z|} 
\sum_{N+M\ge 1}{1\over N !M!}
{\sum}_{(X_{j}),(\Delta_{i})\rightarrow LZ }
\cr
&
{\prod}_{j=1}^{N}\Vert 
K(\lambda, X_{j}) 
\Vert_{h_L,\hat G_{\k,\a}}
{\prod}_{i=1}^{M}
\Vert   \
P (\lambda ,\Delta_{i})
\Vert_{h_L,\hat G_{\k,\a}}
\cr
}$$

$$\Vert\SS(\lambda,K)^\natural\Vert_{h,G_{\k},\AA_{p}}\le 
\sup_\D\sum_{X:L^{-1}\bar{X}^{L}\cap\D\ne\emptyset}\AA_{2+p}(L^{-1}\bar{X}^{L})
\Vert
\bar K_k(X)\Vert_{h_L,\hat G_{\k,\a}}$$

\\Multiply by $\AA_{2+p}(Z)$ on the left and, using 

$$\AA_{2+p}(L^{-1}\bar{X}^{L})\le O(1)\AA_{-3}(X)$$ 

\\(Lemma~1 in [BDH-est]), by 

$$O (1) {\prod}_{j=1}^{N}\AA_{-3}(X_{j})
{\prod}_{i=1}^{M}\AA_{-3}(\Delta _{i})$$

\\on the right. Fix any $\Delta$ and sum both sides over $Z \ni \Delta$. The
spanning tree argument of Lemma~7.1 of [BY] controls the sums over
$N,M, Z, (X_{j}),(\Delta_{i})\rightarrow LZ$ with the result

$$\Vert\SS(\lambda,K)^\natural\Vert_{h,G_{\k},\AA_{p}} 
\le
O(1)\sum_{N\ge 1}O(1)^NL^{3N}
\bigg(
\Vert K (\lambda ) \Vert_{h,\hat G_{\k,\a},\AA} +
\Vert P (\lambda ) \Vert_{h,\hat G_{\k,\a},\AA}
\bigg)^{N}$$

\\The proof of \equ({5.57}) is completed by Lemmas~5.8 and
5.14. When $R=0$ we can replace Lemma~5.14 by Lemmas~5.13 and 5.8
which gives the result with $\eta =0$.

\vskip.3truecm

\\For \equ(5.58) we use \equ({5.19-1}) with $\phi =0$ and replace $h$ by 
$h_{*}$. We estimate the $\zeta$ dependence by the regulator $\tilde G_{\k,\a}$
introduced in \equ({5.2}), to obtain, in the place of
\equ({5.19-2}),

$$\eqalign{
\Vert\SS (\lambda ,K) (Z,0,\z) \Vert_{h_{*}} \le
&2^{|Z|} \tilde G_{\k,\a}(LZ,\z_{L^{-1}},\z_{L^{-1}})
\sum_{N+M\ge 1}{1\over N !M!}
{\sum}_{(X_{j}),(\Delta_{i})\rightarrow LZ }
\cr
&
{\prod}_{j=1}^{N}\Vert 
K(\lambda, X_{j}) 
\Vert_{h_{*},\tilde G_{\k,\a}}
{\prod}_{i=1}^{M}
\Vert   \
P (\lambda ,\Delta_{i})
\Vert_{h_{*},\tilde G_{\k,\a}}
\cr
}
\Eq(5.19-3)$$

\\Then Lemma~5.15 is used to estimate the $\natural$ in $\vert
\SS(\lambda,K)^\natural (Z)\vert_{h_{*}}$, and the rest is as before. End
of proof of Lemma~5.16.

\vglue.3truecm

\\{\it Estimates on relevant parts and flow coefficients 
from the remainder}

\vglue.3truecm

\\Let $({\tilde \a}_{P})$ be  the coefficients $({\tilde \a}_0, 
{\tilde \a}_{2,0},
{\tilde \a}_{2,1}, {\tilde \a}_4)$ defined in \equ(4.57) and \equ({4.62}). 
The flow
coefficients $\xi_{ R}$, $\r_{ R}$ and $\b_{0}$ are given
in \equ({4.64}).

\vglue.3truecm
\\{\it Lemma 5.17

$$\Vert  R^\sharp\Vert_{h,G_{3\k},\AA_{-2}}\le \e^{3/4-\eta}
\Eq(5.65)$$
$$\vert  R^\sharp\vert_{h_{*},\AA_{-3}}\le O(1)\e^{11/4-\eta}
\Eq(5.66)$$
$$\vert{\tilde \a}_{P}\vert_\AA\le O(1)\e^{11/4-\eta}
\Eq(5.67)$$
$$|\b_{0}| \le C\e^{11/4-\eta}
\Eq(5.68)$$
$$|\xi_{ R}|\le C\e^{11/4-\eta}
\Eq(5.69)$$
$$|\r_{ R}|\le C\e^{11/4-\eta}
\Eq(5.70)$$
}

\vglue.3truecm
\\{\it Proof}

\\Recall that ${\tilde \a}_{P}(X)$,  are supported on small
sets. \equ(5.65)follows from the hypothesis \equ({55.2}) and the stability of
the large field regulator $G_{\k}$.  \equ(5.66) follow from the hypotheses
\equ({55.3}) and lemma 5.16 with $n_0 =9$ and $\e$ sufficiently small
depending on L. \equ(5.67)
follows from \equ(5.66), and \equ({4.62}). In fact the dominant contribution
comes by setting ${\tilde V}=0$ because the difference gives additional powers
of $\e$. Then  we have  
$$\Vert {\tilde \a}_{P}\Vert_{\AA} \le O(1) n(P)! h_{*}^{-n(P)}\vert 1_{S}R^\sharp\vert_{h_{*},\AA}$$
where $n(P)$ is the number of fields in the monomial $P$ and $1_{S}$ is the
indicator function on small sets. Now use \equ(5.66) to get\equ(5.67). 
\\
\equ(5.68), \equ(5.69), \equ(5.70) 
follow from
\equ(5.67),and the definitions \equ(4.64), \equ(4.67a) and Wick 
coefficients are $O(1)$. Lemma~5.17 has been proved.

\vglue.3truecm
\\{\it Corollary 5.18

\\For $\e$ sufficiently small
$$\vert g'-\bar g\vert\le \e^{3/2},\ \ \vert\mu'\vert\le C\e^{2-\delta}  \Eq(55.8.1)$$
$$\vert g'- g\vert\le \e^2   \Eq(55.9)$$
}

\\{\it Proof}

\\It is easy to check from the first of the flow equations \equ({4.47}) and the 
definition of $\bar g$ that

$$g'-\bar g=(g-\bar g)(1-L^{2\e}ag) + \xi_{ R}$$

\\and

$$g'- g=(g-\bar g)(-L^{2\e}ag)+ \xi_{ R}$$ 

\\$a=O(\log L)>0$ and the initial Hypothesis implies $g=O(\e)$ so that
for $\e$ sufficiently small $0<1-L^{2\e}ag <1-ag$. The domain of $g$
in the Hypothesis and the bound (5.95) of Lemma~5.17 imply, for $\e$
sufficiently small, that $\xi_{ R}$ is smaller than the other terms,
which gives the two bounds concerning $g'$. The bound on $\mu'$
follows from the second of the flow equations \equ({4.47}), the hypothesis
on $\mu$, and the bound (5.96) on $\rho_{ R}$.

\\The corollary has been proved.  

\vglue.3truecm

\\The following lemma proves the stability of $V$ with respect to 
perturbations by relevant parts $F$ in our model. We state it in the form 
enunciated as the stability hypothesis in Section 4.2, (103), [BDH-est].
This lemma will be very useful for the control of the extraction
formula, as explained in the reference above. 

\\Recall from \equ(4.12) that $F (\lambda) = \lambda^{2}F_{Q}+
\lambda^{3}F_{R}$ and from \equ({33.2}) that (each part of) $F$
decomposes: $F (X) = \sum_{\Delta \subset X}F(X,\Delta)$.

\vglue.3truecm

\\{\it Lemma 5.19

\\For any $R>0$ and $\xi :=R \max(|\lambda^{2}|\e,
|\lambda^{3}|\e^{7/4-\eta})$ sufficiently small,

$$
\Vert e^{-\tilde V_L(\D)-\sum_{X\supset\D} z(X)F(\lambda,
X,\D)}\Vert_{h,G_\k}\le 2^2 \Eq(5.88)$$

\\where $z(X)$ are complex parameters with $|z(X)|\le R$.
}

\vglue.3truecm

\\{\it Proof}

\\First note that Lemma~5.5 still holds if we replace $\tilde V$ by
$\tilde V_L$ provided $\e$ is sufficiently small. We then have

$$
\Vert e^{-\tilde V_L(\D)-\sum_{X\supset\D}z(X)F(X,\D)}\Vert_{h}\le 2\
e^{-g_L/4\int_\D d^3x\phi^4(x)+\sum_{X\supset\D}R\Vert
F(\lambda ,X,\D)\Vert_h}\Eq(55.88) $$

\\Recall that the $F(X,\D)$ are supported on small sets $X$. The proof now 
follows easily from the following

\vglue.3truecm

\\{\it Claim:} For $\e$ sufficiently small

$$\Vert F(\lambda,X,\D)\Vert_h
\le C_{L}\xi\left(\int_\D
d^3x\e\phi^4(x)+ \e^{1/2}\Vert\phi\Vert^2_{\D,1,\s}+1\right)\Eq(5.91)
$$

\\where $\Vert\phi\Vert^2_{\D,1,\s}$ is the norm defined in
\equ({2.2}).

\\{\it Proof of the Claim:} We have
 
$$\Vert F(\lambda,X,\D)\Vert_h\le |\lambda |^{2}\Vert
F_Q(X,\D)\Vert_h+|\lambda |^{3}\Vert F_{ R}(X,\D)\Vert_h \Eq(5.95)$$

\\From \equ(4.32)-\equ(4.36a)

$$\Vert F_Q(X,\D)\Vert_h\le O(1)\e^2\left(\sum_{m=2,4}
\Vert\int_\D d^3x:\phi^m:_C(x)\Vert_h
\sup_{x\in \D}|f^{(m)}_{1,\tilde Q}(x,X,\D)|
+|\tilde Q^{(0,0)}(X,C,v^{(4)})|\right)$$

\\Undoing the Wick ordering produces lower order terms with $O(1)$
coefficients.  Now it is easy to see that for $m=2,4$

$$
\e\Vert\int_\D d^3x:\phi^m:_C(x)\Vert_h\le O(1)
\e\int_\D d^3x\phi^4(x)+O(1)$$

\\By Lemma~5.12,

$$\sup_{x\in \D}|f^{(m)}_{1,\tilde Q}(x,X,\D)|\le C_{L},
\qquad 
|\tilde Q^{(0,0)}(X,C,v^{(4)})|\le  C_{L}$$

\\Therefore

$$|\lambda^{2}|R \ \Vert F_Q(X,\D)\Vert_h
\le C_{L} R |\lambda |^{2}\e \left(\e\int_\D d^3x\phi^4(x)+
1\right)\Eq(5.96)$$

\\Next consider $F_{ R}$, supported on small sets, defined in
\equ(4.57), \equ(4.58). Recall \equ({4.57b}),

$$F_{R}(X,\D,\phi)= \sum_{P}\int_{\Delta } d^3x\ 
\alpha_{P} (X,\Delta ,x)P (\phi (x),\dpr\phi (x))
$$

\\By Lemma~5.17 and \equ(4.67a)

$$
|\a_P(X,\Delta ,x)|\le C_{L} \e^{11/4-\eta}$$

\\so that

$$\eqalign{
|\lambda |^{3}\vert z(X)\vert \Vert F_{ R}(X,\D)\Vert_h
\le &C_{L} R|\lambda |^{3}\e^{11/4-\eta}\sum_{P}
\int_{\Delta } d^3x\ \Vert
P (\phi (x),\dpr\phi (x))
\Vert_h\cr
\le &C_{L} R|\lambda |^{3}\e^{7/4-\eta}\bigg( 
\e\int_\D d^3x \, \phi^4(x)+\e^{1/2}\Vert\phi\Vert^2_{\D,1,\s}+1
\bigg)\cr}$$

\\The claim follows by combining this with \equ({5.96}).

\\Note that in the above inequality the Sobolev norm arises only when
estimating the $j$-term corresponding to $\phi\dpr_\m\phi$. For this
we can bound $\vert\phi\nabla\phi\vert \le 1/2(\vert\phi\vert^2
+\vert\nabla\phi\vert^2)$ and then use the Sobolev embedding
inequality.  End of proof of Lemma.

\vglue.3truecm

\\{\it Lemma 5.20

\\For any $R>0$ and $\xi :=R \max(|\lambda^{2}|\e^{2},
|\lambda^{3}|\e^{11/4-\eta})$ sufficiently small,

$$
\vert e^{-\tilde V_L(\D)-\sum_{X\supset\D}
z(X)F(\lambda, X,\D)}\vert_{h_{*}}\le 2^2
\Eq(5.88.1)$$

\\where $z(X)$ are complex parameters with $|z(X)|\le R$.
}

\vglue.3truecm

\\{\it Proof}

\\This is the same as the last proof except that we can use the estimate

$$\vert F(\lambda,X,\D)\vert_{h_{*}}
\le C_{L}\xi
$$

\\in place of \equ({5.91}).  (No need to ensure $R\vert F(\lambda
,X,\D)\vert_{h_{*}}$ is smaller than $\e \phi^{4}$ because stability
away from $\phi =0$ is not an issue with the kernel norm). End of
proof of Lemma.

\vglue.3truecm
\\Recall \equ({4.188})

$$
R_{\rm main} 
= {1\over 2\pi i} 
\oint_{\gamma} {d\lambda \over \lambda^{4}}
{\cal E} \bigg( 
\SS(\lambda ,Qe^{-V})^{\natural},F_{Q} (\lambda ) 
\bigg)
\Eq(5.lem35-0)
$$

\vglue.3truecm
\\{\it Lemma 5.21

$$\Vert R_{\rm main}\Vert_{h,G_{\k},\AA}\le C_{L}\e^{3/4}
\Eq(5.lem35-1)$$

$$\vert R_{\rm main}\vert_{h_{*},\AA} \le C_{L}\e^{3-3\d/2}
\Eq(5.lem35-2)$$

}

\vglue.3truecm

\\{\it Proof}

\\Let

$$J(\lambda) = \SS(\lambda ,Qe^{-V})^{\natural}$$

\\Suppose that $F (\lambda )$ splits, $F (\lambda)= F_{0}
(\lambda)+F_{1} (\lambda)$, into a field independent part $F_{0}$ and
$F_{1}$ satisfies stability as in \equ({5.88}).  According to Theorem
6 on page 780 of [BDH-est],

$$
\Vert {\cal E}(J(\lambda),F_{0},F_{1})  \Vert_{h,G_{\k},\AA }
\le
O (1) \bigg(
\Vert J(\lambda) \Vert_{h,G_{\k},\AA_{2} } + \Vert f  \Vert_{\AA_{4}}
\bigg)
\Eq({5.lem35-3})$$

\\provided the norms on the right hand side are less than a small
constant $R^{-1}=O (1)$. In the above 
$ |f(X)| \le 2|z(X)|^{-1}$ where the $z(X)$
are the complex parameters introduced in Lemma~5.19. The $f(X)$ are 
supported on small sets. We choose $ |\lambda | = c_{L}\e^{-1/4}$.
By Lemma~5.19 we have stability \equ({5.88}) if $\e$ is small.
Therefore \equ({5.lem35-3}) holds and by combining it with
Lemma~5.16,

$$
\Vert {\cal E}(J(\lambda),F_{0},F_{1})  \Vert_{h,G_{\k},\AA }
\le q +O (R^{-1}) \le O (1)
$$

\\\equ({5.lem35-1}) follows by choosing the contour $\gamma $ in 
\equ({5.lem35-0}) to be a circle of radius $c_{L}\e^{-1/4}$.

\hskip.3truecm

\\The proof of \equ({5.lem35-2}) follows the same steps but with
contour $\gamma$ being a circle of radius $c_{L}\e^{-1+\d/2}$
chosen so that $\xi$ in Lemma~5.20 is small and the hypothesis of
Lemma 5.16 is satisfied. End of proof.

\vskip.3truecm

\\Recall from \equ({4.21}) that

$$
R_{3} = 
{1\over 2\pi i} 
\oint_{\gamma} {d\lambda \over \lambda^{4} (\lambda -1)}
{\cal E} (\SS (\lambda,K)^{\natural},F (\lambda
))
\Eq(5.lem36-3)$$

\vglue.3truecm
\\{\it Lemma 5.22

$$\Vert R_{3}\Vert_{h,G_{\k},\AA}\le C_{L}\e^{1-4\eta /3}
\Eq(5.lem36-1)$$

$$\vert R_{3}\vert_{h_{*},\AA} \le C_{L}\e^{11/3-4\eta /3}
\Eq(5.lem36-2)$$

}

\vglue.3truecm

\\{\it Proof}

\\This proof follows the same steps as the proof of Lemma~5.21 with
contours $|\lambda|= c_{L}\e^{1/4-\eta/3}$ and $|\lambda|=
c_{L}\e^{11/12-\eta/3}$.  End of proof.

\vglue.3truecm

\\{\it Lemma 5.23

\\$R_{4}$ as defined in \equ(4.22) satisfies

$$
\eqalign{
&\Vert R_{4}  \Vert_{h,G_{\kappa},\AA} 
\le C \e^{3/2}\cr
&\vert R_{4}  \vert_{h_{*},\AA} 
\le C \e^{3}\cr
}
$$
}

\vglue.3truecm

\\{\it Proof}

\\From \equ(4.22)

$$
R_{4} = 
\bigg(
e^{-V'} - e^{-\tilde{V}_{L}}
\bigg)
Q(C,{\bf w}',g') + 
e^{-\tilde{V}_{L}}
\bigg(
Q(C,{\bf w}',g') - Q(C,{\bf w}',g_L)
\bigg)
$$

First we observe from \equ(4.47A), Lemma~5.17 and Lemma~5.9 the
following bounds

$$\vert g'-g_L\vert \le C\e^2  \Eq(5.lemm37-1)$$ 
$$\vert \m'-\m_L\vert \le C\e^2  \Eq(5.lemm37-2)$$
$$\Vert {\bf w}' \Vert \le c/4  \Eq(5.lemm37-3)$$

\\We estimate in turn the two terms in the expression for $R_4$
above. Because of Q each term is supported on small sets.

\vglue.3truecm

\\We write the first term as

$$\bigg(
e^{-V'} - e^{-\tilde{V}_{L}}
\bigg)
Q(C,{\bf w}',g')=Q(C,{\bf w}',g')e^{-{1\over 2}V'}\int_0^1 ds (\tilde{V}_{L} -
V')e^{-({1\over 2}-s)V'-s{V}_{L}}$$

\\Then we bound

$$\Vert \bigg(e^{-V'} - e^{-\tilde{V}_{L}}\bigg)Q(C,{\bf
w}',g')\Vert_{h,G_{\k}} \le \Vert Q(C,{\bf w}',g')e^{-{1\over
2}V'}\Vert_{h,G_{\k} } \int_0^1 ds \Vert \tilde{V}_{L}
-V'\Vert_{h}\Vert e^{-({1\over 2}-s)V'}\Vert_{h}\Vert
e^{-s{V}_{L}}\Vert_{h} \Eq(5.lemm37-4) $$

\\Using \equ(5.lemm37-1) and \equ(5.lemm37-2) we can bound 

$$\Vert \tilde{V}_{L}(X) -V'(X)\Vert_{h}\le C{\e\over \g}
e^{O(1)\g\e\int_{X}d^{3}x \phi^{4}(x)}\Eq(5.lemm37-5)   $$
for any $\g=O(1) >0$.

\\By Lemma 5.5 we can bound

$$\Vert e^{-({1\over 2}-s)V'(X)}\Vert_{h} \le 2^{|X|}
e^{-({1\over 2}-s){g'\over 4}\int_{X}d^{3}x \phi^{4}(x)}\Eq(5.lemm37-6) $$
$$\Vert e^{-s){V}_{L}(X)}\Vert_{h} \le 2^{|X|}
e^{-s{g'\over 4}\int_{X}d^{3}x \phi^{4}(x)}\Eq(5.lemm37-7) $$

\\We now plug into \equ(5.lemm37-4) the bounds \equ(5.lemm37-6) and
\equ(5.lemm37-7). We then write the $s$-integration in \equ(5.lemm37-4) 
as the union of the intervals $[0,{1\over 4}$ and ${1\over 4},1]$.
In the first interval we insert the bound \equ(5.lemm37-5) with $\g$
replaced by $({1\over 2}-s)\g$. In the second interval we insert the
same bound with  $\g$ replaced by $s\g$ and in both cases take $\g=O(1)$
sufficiently small. Then the $s$-integral factor in \equ(5.lemm37-4)
is bounded by $C\e$. On the other hand the first factor in \equ(5.lemm37-4)
is bounded by $O(1)\e^{1\over 2}$ by virtue of Lemma~5.10. (The factor of
$1\over 2$ in the exponent does not make a difference). Putting these bounds 
together and recalling that the $Q$ are supported on small sets we
obtain

$$\Vert \bigg(e^{-V'} - e^{-\tilde{V}_{L}}\bigg)Q(C,{\bf
w}',g')\Vert_{h,G_{\k},\AA} \le C\e^{3\over 2} \Eq(5.lemm37-8)
$$

\\We can now easily bound the second term in the expression for $R_4$ by noting 
that

$$\vert g'^{2}-g_{L}^2\vert \le C\e^3$$

\\and then using Lemma~5.10. We again get the bound 

$$\Vert e^{-\tilde{V}_{L}}
\bigg(
Q(C,{\bf w}',g') - Q(C,{\bf w}',g_L)
\bigg)\Vert_{h,G_{\k},\AA} \le  C\e^{3\over 2} \Eq(5.lemm37-9) $$ 

\\Adding together \equ(5.lemm37-8) and \equ(5.lemm37-9) finishes the proof
of the first bound in the Lemma. The second bound is easy to prove since
all derivatives in the $h_*$ norm are at $\phi =0$. End of proof of 
Lemma~5.23.

\vglue.3truecm
\\{\it Lemma 5.24

\\Let $X$ be a small set and let $J$ be normalized as in
\equ({4.61}). Then we have

$$
\vert D^2J(X,0;f^{\times 2}_{L^{-1}})\vert\le
O(1)L^{-(7-\e)/2}\Vert
D^2J(X,0)\Vert\prod_{j=1}^2\Vert f_j\Vert_{C^2(L^{-1}X)}\Eq(5.79)$$
$$\vert D^4J(X,0;f^{\times 4}_{L^{-1}})\vert\le
O(1)L^{-(4-\e)}\Vert
D^4J(X,0)\Vert\prod_{j=1}^4\Vert f_j\Vert_{C^2(L^{-1}X)}\Eq(5.80)$$
}

\vglue.3truecm

\\{\it Proof: }  See Lemma~15 [BDH-est].

\vglue.3truecm

\\{\it Corollary 5.25

\\Let $Y=L^{-1}X$ where $X$ is a small set, $Z=L^{-1}\bar{X}^{L}$
and let $J$ be normalized as in \equ({4.61}). Then

$$
|J_{L} (Y)|_{h} \le O (1) L^{-(7-\e)/2}|J(X)|_{h}
\Eq(5.lem39-1)
$$

$$
\Vert 
J_{L} (Y) e^{-\tilde{V}_{L} (Z\setminus Y)}
\Vert_{h,G_{\k}} \le O (1) L^{-(7-\e)/2}
\bigg[
|J(X)|_{h} + \Vert J(X)\Vert_{h,G_{3\k}} 
\bigg]
\Eq(5.lem39-2)
$$
}

\vglue.3truecm

\\{\it Proof}

\vglue.3truecm

\\\equ({5.lem39-1}) follows immediately from
Lemma~5.24. For \equ({5.lem39-2}) we write

$$
\Vert J_{L} (Y,\phi) e^{-\tilde{V}_{L} (Z\setminus Y,\phi)}\Vert_{h}
\le
\Vert J_{L} (Y,\phi) \Vert_{h}
\Vert e^{-\tilde{V}_{L} (Z\setminus Y,\phi)}\Vert_{h}
$$

\\and use Lemmas~5.5 and Lemma~5.15,

$$
\le O (1)G_\k(Z,\phi)\bigg[
|J_{L}(Y)|_{h} + 
L^{-n_{0}[\phi]}\Vert J_{L} (Y)\Vert_{L^{[\phi]}h,G_{\kappa}} 
\bigg]
$$

\\followed by \equ({5.lem39-1}), and rewriting the second term by
moving the scaling from $J$ to the norm,

$$
\le
O (1)G_\k(Z,\phi)\bigg[
L^{-(7-\e)/2}|J(X)|_{h} + 
L^{-n_{0}[\phi]}\Vert J(X)\Vert_{h,G_{3\kappa}} 
\bigg]
$$

\\\equ({5.lem39-2}) follows by multiplying both sides by
$G_\k^{-1}(Z,\phi)$ and taking the supremum over $\phi$.  End of
proof.

\vglue.3truecm

\\{\it Lemma 5.26

$$\Vert\tilde F_{R}e^{-\tilde V}\Vert
_{h,G_{\k},\AA}\le O(1)\e^{7/4-\eta}
\Eq(5.71)$$

$$\vert\tilde F_{R}e^{-\tilde V}\vert
_{h_{*},\AA}\le O (1)\e^{11/4-\eta}
\Eq(5.72)$$

\\and $J= R^{\sharp}-\tilde F_{R}e^{-\tilde{V}}$ satisfies on small
sets the bounds}

$$\Vert J\Vert_{h,G_{3\k},\AA}\le O(1)\e^{{3\over 4}-\eta} \Eq(5.722a) $$
$$\vert J\vert_{h_{*},\AA} \le O(1)\e^{{11\over 4}-\eta}$$

\vglue.3truecm
\\{\it Proof}

\\First we prove \equ({5.71}).

\\Take the definition of $\tilde F_{R}$ given in \equ(4.57) and
\equ({4.59}). $\tilde F_{R}$ is supported on small sets. We have

$$\Vert {\tilde F}_{ R}(X,\phi)\Vert_h
\le \sum_{P} \vert {\tilde \a}_{P}(X)\vert
\int_{X } d^3x\ \Vert
P (\phi (x),\dpr\phi (x))\Vert_h$$
$$\le O(1)\sum_{P}\vert {\tilde \a}_{P}(X)\vert \e^{-1}
\bigg( 
\e\int_{X} d^3x \, \phi^4(x)+\e^{1/2}\Vert\phi\Vert^2_{X,1,\s}+1
\bigg)$$
$$\le O(1)\sum_{P}\vert {\tilde \a}_{P}(X)\vert \e^{-1}G_\k(X,\phi)e^{\g g\int_Xd^3y\phi^4(y)}
$$
for any $\g =O(1)>0$. Hence, using Lemma 5.5
$$\Vert\tilde F_{R}(X\phi)e^{-\tilde V(X\phi)}\Vert_h
\le\Vert\tilde F_{R}(X\phi)\Vert
_h\Vert e^{-\tilde V(X\phi)}\Vert_h \le O(1)\sum_{P}2^{|X|}\vert{\tilde \a}_{P}
(X)\vert \e^{-1}G_\k(X,\phi)    $$
We thus obtain ( remembering that ${\tilde \a}_{P}$ are supported on small
sets) on using \equ(5.67) 
$$\Vert\tilde F_{R}(X\phi)e^{-\tilde V(X\phi)}\Vert_{h,G_{\k},\AA}
\le O(1)\e^{-1}\sum_{P}\Vert \a_{P}\Vert_{\AA} \le O(1)\e^{7/4 -\eta}$$
This proves \equ({5.71}).

\\Now we turn to the proof of \equ({5.72}).

\\As observed in the proof of Lemma~5.17, for $\e$ sufficiently small 
(depending on L),
$$\vert{\tilde \a}_{P}\vert_{\AA} \le n(P)! h_{*}^{-n(P)}\vert 1_{S}R^\sharp\vert_{h_{*},\AA} \le O(1) h_{*}^{-n(P)} \e^{11/4-\eta}$$
We have from  the definition of$\tilde F_{R}$ given in \equ(4.57)
$$\vert \tilde F_{R}(X)\vert_{h_{*}} \le O(1)\sum_{P} \vert {\tilde \a}_{P}(X)\vert h_{*}^{n_P}$$  

\\whence
$$\vert \tilde F_{R}\vert_{h_{*},\AA} \le \sum_{P}\vert{\tilde \a}_{P}\vert_{\AA}h_{*}^{n_P}\le O(1)\e^{11/4-\eta}$$ 
which proves \equ({5.72}). 

\vglue.3truecm

\\To get these bounds for $J= R^{\sharp}-\tilde F_{R}e^{-\tilde{V}}$
we apply \equ({5.71}) and \equ({5.72}) to the $\tilde
F_{R}e^{-\tilde{V}}$ part. We bound $R^{\sharp}$ by Lemma~5.17. End of proof.

\vglue.3truecm

\\{\it Corollary 5.27}
$$\Vert R_{\rm linear}\Vert
_{h,G_{\k},\AA}\le O(1)L^{- (1-\e)/2}\e^{3/4-\eta}
\Eq(5.73)$$

$$\vert R_{\rm linear}\vert
_{h_{*},\AA}\le O(1)L^{- (1-\e)/2}\e^{11/4-\eta}
\Eq(5.74)$$

\vglue.3truecm

\\{\it Proof}

\vglue.3truecm

\\We recall from \equ({4.59}) that

$$
J= R^{\sharp}-\tilde F_{R}e^{-\tilde{V}}
$$

\\is normalized. Let $1_{\rm s.s}(X)$ be the indicator function of the
event that $X$ is small.  Referring to \equ({4.59}), the first term
in $R_{\rm linear} (Z)$ is

$$
R_{\rm
linear, s.s} (Z) := \sum_{X: L^{-1}\bar{X}^{L}=Z}
e^{-\tilde{V}_{L} (Z\setminus Y)}1_{\rm s.s}(X)J_{L} (Y)
\Eq(4.59b3)
$$

\\where $Y=L^{-1}X$.  By Corollary~5.25 this is bounded in $h,G_{\k}$
norm by

$$
O (1)L^{(7-\e)/2}\sum_{X:L^{-1}\bar{X}^{L}=Z}
1_{\rm s.s}(X)      
\bigg[
|J(X)|_{h} + \Vert J(X)\Vert_{h,G_{3\k}} 
\bigg]
\Eq(4.59b1)
$$

\\Multiply both sides by $\AA(Z)$, using $\AA(Z) \le \AA(X)$ on
the right hand side.  Then sum over $Z$ to get the $\AA$ norm and use
the bounds on $J$ in Lemma~5.17 and Lemma~5.26 to obtain

$$
\Vert R_{\rm linear, s.s} \Vert_{h,G_{\k},\AA} 
\le O(1)L^{- (1-\e)/2}\e^{3/4-\eta}
$$

\\where we used an argument on page 790 of [BDH-est] to control the
$\sum_{X: L^{-1}\bar{X}^{L}=Z}$ by $L^{D=3}$ times the sum over $X$ in the
definition of $\AA$ norm. Similarly the bound on the kernel norm by $
O(1)L^{- (1-\e)/2}\e^{11/4-\eta}$ comes from the kernel norm bound in
Lemma~5.17, Lemma~5.26 and Lemma~5.15.

\vglue.3truecm

\\Let $1_{\rm l.s}(X)$ be the large set indicator function. The second
term in $R_{\rm linear} (Z)$ is

$$
R_{\rm
linear, s.s} (Z) := \sum_{X: L^{-1}\bar{X}^{L}=Z}
e^{-\tilde{V}_{L} (Z\setminus L^{-1}X)} 1_{\rm l.s}(X)R^{\natural}_{L} (L^{-1}X)
\Eq(4.59b2)
$$

\\where we have used $J (X)=R^{\sharp}(X)$ because the subtraction is
supported on small sets.  This is bounded in the same way as above
using Lemma~5.17 except that the necessary $L^{- (1-\e)/2}$ is
obtained for a different reason: For large sets, by \equ({2.6}), $\AA
(Z) \le c_{p} L^{-4}\AA_{-p}(X)$, where $c_{p}= O(1)$.  The Corollary
is proved.

\vglue.3truecm

\\{\it Proof of Theorem~1 Concluded}

\vglue.2truecm

\\From \equ({4.16})-\equ({4.23}), $R$ is the sum of $R_{i}$ where
$i=1,2,3,4$.  By Lemmas~5.21, 5.22, 5.23 and 5.27 with $L$ large
and $\e$ small depending on $L$, the sum satisfies bounds \equ({55.7})
and \equ({55.8}).

\vglue.5truecm

\\{\bf 6. STABLE MANIFOLD AND CONVERGENCE TO NON-GAUSSIAN
FIXED POINT} 
\numsec=6\numfor=1

\vskip0.5truecm
\\{\it 6.1 }
 
\\Let $\bar g$ be the approximate fixed point of the $g$ flow
given by \equ(55.4). Let us define
$$\tilde g=g-\bar g\Eq(6.1)$$
and 
$$u=(\tilde g,\m,R,{\bf w})\Eq(6.2)$$
Then the RG iteration given by \equ(4.47A), \equ({4.23}), \equ(4.54)
can be written as
$$u'=f(u)\Eq(6.3)$$
with components
$$\tilde g'=f_g(u)=(2-L^\e)\tilde g +\tilde\xi(u)\Eq(6.4)$$
$$\m'=f_\m(u)=L^{3+\e\over 2}\m+\tilde\r(u)\Eq(6.5)$$
$$R'=r_R(u)=U(u)\Eq(6.6)$$
$${\bf w}'=f_{\bf w}(u)={\bf v}+{\bf w}_L\Eq(6.7)$$
with initial
$$u_0=(\tilde g_{0},\m_{0},0,0)\Eq(6.8)$$ 
Here
$$\tilde\xi(u)=-L^{2\e}a\tilde g^2 +\xi(u)\Eq(6.9)$$
$$\tilde \r(u)=-L^{2\e}b(g+\tilde g)^2 +\r(u)\Eq(6.10)$$
Note that the ${\bf w}$ flow is autonomous and solved by 
\equ(5.19). In Lemma~5.9 it was proved that the ${\bf w}$
tends to the fixed point ${\bf w}_*$ in an appropriate norm.

\\Let $E$ be the Banach space consisting of elements $u$ with the (box) norm
$$\Vert u\Vert=\sup(\e^{-3/2}|\tilde g|,\e^{-(2-\d)}|\m|,
\e^{-(11/4-\eta)}|||R|||,c^{-1}\Vert{\bf w}\Vert)\Eq(6.11)$$
Here
$$|||R|||=\sup(\e^2\Vert R\Vert_{h,G,\AA},\vert R\vert_{h_{*},\AA})\Eq(6.12)$$
and
$$\Vert{\bf w}\Vert=\sup_p\Vert w^{(p)}\Vert_p$$
The $\Vert\cdot\Vert_p$ and $\Vert {\bf w}\Vert$ norms were defined in 
\equ(5.20) and the constant $c$ is that of Lemma~5.9.

\\Let $B(r)\subset E$ be the closed ball of radius $r$,
centered at the origin:
$$B(r)=\{u\in E: \Vert u\Vert\le r\}\Eq(6.13)$$
Let $\cal D$ be the domain of $(g,\m,R)$ specified in
\equ(55.1)-\equ(55.3) of the hypothesis
stated at the beginning of Section 5. Then we have
$$u\in B(1)\ \Rightarrow\ (g,\m,R)\in\cal D\Eq(6.14)$$
and then Theorem 1 of Section 5 holds.

\\Theorem 1, together with the autonomous Lemma~5.9, shows that
the ball $B(1)$ would be stable under the RG flow $f$, except
for the unstable direction $\m$ (as evident from \equ(6.5)).
The initial unstable parameter $\m_0$ will have to be fine tuned
to a critical function $\m_0=\m_{c}(g_0)$ which would
determine the critical or stable submanifold in which 
we expect to get a contraction to a fixed point.
As observed in [BDH-eps] this is part of the stable
manifold theorem in the theory of hyperbolic dynamical
systems [S]. In the following we put ourselves in this framework
(see appendix 2, Chapter 5 of [S] and Section 5 of
[BDH-eps]).

\\Suppose $u\in B(1)$. Then from theorem 1 of Section 5 we have
for $\e>0$ sufficiently small (depending on $L$), which implies
in particular $L^\e=O(1)$,
$$\eqalign{
|\xi(u)|\le &O(1)\e^{11/4-\eta}\cr
|\r(u)|\le &O(1) L^{3/2}\e^{11/4-\eta}\cr
|||U(u)|||\le &L^{-1/4}\e^{11/4-\eta}\cr}\Eq(6.15)$$
>From \equ(6.9), \equ(6.10) we have by virtue of \equ(6.15) 
and the estimates $a=O(\log L)$, $b=O(L^{3/2}$,    see Lemma~5.12, for
$u\in B(1)$:

$$\eqalign{
|\tilde\xi(u)|\le &O(1)\e^{11/4-\eta}\cr
|\tilde\r(u)|\le &\e^{2-\d}\cr}\Eq(6.16)$$
$\tilde\xi,\tilde\r,U,{\bf w}$ satisfy the following Lipshitz
bounds in $B(1/4)\subset B(1)$

\vglue.3truecm
\\{\it Lemma 6.1}

\\Let $u,u'\in B(1/4)$. Then we have the Lipshitz
bounds:

$$\eqalign{
|\tilde\xi(u)-\tilde\xi(u')|\le &O(1)\e^{11/4-\eta}\Vert u-u'\Vert\cr
|\tilde\r(u)-\tilde\r(u')|\le &
O(L^{3/2})\e^{5/2}\Vert u-u'\Vert\cr
|||U(u)-U(u')|||\le &O(1)L^{-1/4}\e^{11/4-\eta}\Vert u-u'\Vert\cr
\Vert f_{\bf w}(u)- f_{\bf w}(u')\Vert\le &cL^{-1/4}
\Vert u-u'\Vert\cr}
\Eq(6.17)$$

\\{\it Proof}

\\We shall use the fact that $\xi,\r,U$ are analytic in $B(1)$.
This follows from the algebraic operations in Section 4
together with the analyticity of the extraction map.

\\Let $\D u=u-u'$. Then
$$U(u)-U(u')=\int_0^1dt{\dpr\over\dpr t}U(u+t\D u)$$
By the Cauchy integral formula
$${\dpr\over\dpr t}U(u+t\D u)=
{1\over 2\pi i}\oint_\g dz {U(u+z\D u)\over (z-t)^2}$$
here $0\le t\le 1$ and $\g$ is the closed contour
$$\g:\ z-t=r e^{i\th},\ r=1/4\Vert\D u\Vert^{-1}$$
Note that
$$u+z\D u=u+t\D u +{1\over 4}{\D u\over \Vert\D u\Vert}e^{i\th}$$
Hence
$$\Vert u+z\D u\Vert\le {1\over 4}+{1\over 2}+{1\over 4}=1$$
So that for $z\in \g$, $u+z\D u\in B(1)$, and hence from
\equ(6.15)
$$|||U(u+z\D u)|||\le L^{-1/4}\e^{11/4-\eta}
$$
so that
$$\sup_{0\le t\le 1}|||{\dpr\over\dpr t}U(u+t\D u)|||
\le O(1) L^{-1/4}\e^{11/4-\eta}\Vert\D u\Vert$$
and thus
$$|||U(u)-U(u')|||\le O(1)L^{-1/4}\e^{11/4-\eta}\Vert u-u'\Vert$$
This proves the last inequality of \equ(6.17).

\\On using \equ(6.16) we have in the same way
$$|\tilde\xi(u)-\tilde\xi(u')|\le O(1)\e^{11/4-\eta}\Vert u-u'\Vert$$
To get the Lipshitz bound for $\tilde \r$ we first use
$$|\r(u)-\r(u')|\le O(1)L^{3/2}\e^{11/4-\eta}\Vert u-u'\Vert$$
Then from \equ(6.10)
$$\eqalign{|\tilde\r(u)-\tilde\r(u')|\le &O( L^{3/2})|\tilde g'+\tilde g+
2\bar g||\tilde g'-\tilde g| +|\r(u)-\r(u')| \cr
& \le \left(O(L^{3/2})O(\e)e^{3/2}+O(1)L^{3/2}\e^{11/4-\eta}\right)
\Vert u-u'\Vert \cr
&\le O(L^{3/2})\e^{5/2}\Vert u-u'\Vert}$$
Finally from \equ(6.7) and the definition of the norms in \equ(5.20)
we have
$$\Vert f_{\bf w}(u)- f_{\bf w}(u')\Vert\le \sup_{1\le p\le 3}
{\rm ess\, sup}_x\left(|x|^{6p+1\over 4}|w_L^{(p)}(x)-w_L^{(p)}\phantom{,}\!\!
'(x)|\right)$$
Now $w_L^{(p)}(x)=L^2[\phi]w^{(p)}(Lx)$. We then get easily
$$\Vert f_{\bf w}(u)- f_{\bf w}(u')\Vert\le L^{-1/4}\Vert {\bf w}-{\bf w}'\Vert
\le c L^{-1/4}\Vert u-u'\Vert$$
and we are done. Lemma 6.1 has been proved.
\vglue.3truecm
\\Consider now the RG flow \equ(6.3):
$$u_k=f(u_{k-1})$$
with initial condition
$$u_0=(\tilde g_0,\m_0,0,0)\quad \tilde g_0=g_0-\bar g$$
 
\vglue.3truecm
\\{\it Theorem 6.2}

\\There exists  $\m_0$ such that for 
$u_0\in B(1/32)$, $u_k=f(u_{k-1})\in B(1/4)$ for all $k\ge 1$.
\vglue.3truecm
\\{\it Remark} 

\\The following proof of existence of global solutions is a textbook
argument in the theory of dynamical systems adapted to the present
context.  \vglue.3truecm

\\{\it Proof}

\\From the flows \equ(6.4), \equ(6.5) we easily derive after
$n$ steps of the RG
$$\tilde g_k=(2-L^\e)^k\tilde g_0 +
\sum_{j=0}^{k-1}(2-L^\e)^{k-1-j}\tilde\xi(u_j),\quad 1\le k\le n$$
$$\m_k=L^{-{3+\e\over 2}(n-k)}\m_n-
\sum_{j=k}^{n-1}L^{-{3+\e\over 2}(j+1-k)}\tilde\r(u_j),
\quad 0\le k\le n-1$$
Let us fix $\m_n=\m_f$ and take $n\rightarrow\io$. We have
$$\tilde g_k=(2-L^\e)^k\tilde g_0 +
\sum_{j=0}^{k-1}(2-L^\e)^{k-1-j}\tilde\xi(u_j),\quad k\ge 1\Eq(6.18)$$
$$\m_k=-
\sum_{j=k}^{\io}L^{-{3+\e\over 2}(j+1-k)}\tilde\r(u_j),
\quad k\ge 0\Eq(6.19)$$
together with
$$R_k=U(u_{k-1}), \quad k\ge 1\Eq(6.20)$$
We can take the autonomous ${\bf w}$ flow, given by \equ(6.7),
as solved by \equ(5.19) and need no longer consider it as a flow variable.

\\Note that for $\e$ sufficiently small (depending on $L$)
$$0<2-L^{\e}<1\Eq(6.21)$$
Then for $u_j\in B(1)$ the infinite sum of \equ(6.19)
converges by \equ(6.21) and \equ(6.16). So $\m_0$ has now
been determined provided \equ(6.18)-\equ(6.20) has a solution.

\\It is easy to verify that any solution of \equ(6.18)-
\equ(6.19), together with the autonomous ${\bf w}$ flow, is
a solution of the RG flow
$$u_k=f(u_{k-1})$$
Now write \equ(6.18)-\equ(6.20) in the form
$$u_k=F_k({\bf u})\Eq(6.22)$$
where ${\bf u}=(u_0, u_1, u_2,...)$ and $F_k$ has components
$(F_k^{(g)},F_k^{(\m)},F_k^{(R)})$ given by the r.h.s. of
\equ(6.18), \equ(6.19)
and \equ(6.20) respectively.

\\If we write
$${\bf F}({\bf u})=(F_0({\bf u}), F_1({\bf u}),...)$$
then \equ(6.22) can be  written as a fixed point equation
$${\bf u}={\bf F}({\bf u})\Eq(6.23)$$
Consider the Banach space ${\bf E}$ of sequences
${\bf u}=(u_0, u_1, u_2,...)$ with norm
$$\Vert{\bf u}\Vert=\sup_{k\ge 0}\Vert u_k\Vert\Eq(6.24)$$
and the closed ball ${\bf B}(r)\subset{\bf E}$
$${\bf B}(r)=\{{\bf u}:\Vert{\bf u}\Vert\le r\}\Eq(6.25)$$
We shall seek a solution of \equ(6.23) in the closed ball 
${\bf B}(1/4)$ with initial data
$u_0=(\tilde g_0,\m_0,0,0)$  in ${B}(1/32)$ and $\tilde g_0$ held fixed.
 
We shall need

\vglue.3truecm
\\{\it Lemma 6.3}

$${\bf u}\in {\bf B}(1/32)\Rightarrow{\bf F}({\bf u})
\in {\bf B}(1/16)\Eq(6.26)$$
Moreover, for ${\bf u},{\bf u}'\in {\bf B}(1/4)$
$$\Vert{\bf F}({\bf u})-{\bf F}({\bf u}')\Vert\le
{1\over 2}\Vert {\bf u}-{\bf u}'\Vert\Eq(6.27)$$
We postpone the proof of this lemma. Given the above lemma 6.3
the proof of theorem 6.2 now follows easily.
To this end consider a sequence
${\bf u}^{(n)}, n=1,2,...$, defined by
$${\bf u}^{(n)}={\bf F}({\bf u}^{(n-1)})\Eq(6.28)$$

$${\bf u}^{(0)}\in {\bf B}(1/32)\Eq(6.29)$$

\\{\it Claim}: ${\bf u}^{(n)}\in {\bf B}(1/4)$ for all $n$.

\\To prove the claim first observe that \equ(6.26) and
\equ(6.29) imply
$$\Vert{\bf F}({\bf u}^{(0)})-{\bf u}^{(0)}\Vert\le{3\over 32}$$
Make the inductive hypothesis
${\bf u}^{(j)}\in {\bf B}(1/4),\quad j=0,1,...,n-1$
and this is clearly satisfied for $j=0$. Now using the Lipshitz bound
\equ(6.27) we get
$$\Vert{\bf u}^{(n)}-{\bf u}^{(n-1)}\Vert=\Vert{\bf F}({\bf u}^{(n-1)})-
{\bf F}({\bf u}^{(n-2)})\Vert\le{1\over 2^{n-1}}
\Vert{\bf u}^{(1)}-{\bf u}^{(0)}\Vert\le$$
$$\le \Vert{\bf F}({\bf u}^{(0)})-{\bf u}^{(0)}\Vert
\le{1\over 2^{n-1}}{3\over 32}$$
Write
$${\bf u}^{(n)}={\bf u}^{(0)}+\sum_{j=1}^n({\bf u}^{(j)}-
{\bf u}^{(j-1)})$$
Then
$$\Vert{\bf u}^{(n)}\Vert\le{1\over 32}
\left(1+3\sum_{j=1}^n{1\over 2^{j-1}}\right)
\le {7\over 32}< {1\over 4}$$
and the claim has been proved.

\\By virtue of the claim and $\Vert{\bf u}^{(n)}-{\bf u}^{(n-1)}\Vert
\rightarrow 0$ as $n\rightarrow\io$ we have
${\bf u}^{(n)}\rightarrow{\bf u}$ in ${\bf B}(1/4)$ and this ${\bf u}$
satisfies
$${\bf u}={\bf F}({\bf u})$$
We are done. Thus it only remains to prove lemma 6.3 to complete the proof
of theorem 6.2.

\vglue.3truecm
\\{\it Proof of lemma 6.3}

\\First we prove \equ(6.26), and thus take ${\bf u}\in {\bf B}(1/32)$.
>From \equ(6.18) and the estimates in \equ(6.16) we have
$$\e^{-3/2}|F_k^{(g)}({\bf u})|\le 
(2-L^\e){1\over 32}+O(1) \e^{5/4-\eta}
\sum_{j=0}^{k-1}(2-L^\e)^{k-1-j}\le$$
$$\le{1\over 32}+O(1) {\e^{5/4-\eta}\over L^\e-1}\le
{1\over 32}+O(1) {\e^{1/4-\eta}\over \log L}\le 
{1\over 16}$$
since $\eta < {1\over 4}$ and $\e$ is sufficiently small.

\\Similarly from \equ(6.19) and \equ(6.16) we have
$$\e^{-(2-\d)}|F_k^{(\m)}({\bf u})|\le \sum_{j=k}^{\io}
L^{-{3+\e\over 2}(j+1-k)}\le L^{-{3+\e\over 2}}(1-L^{-{3+\e\over 2}})\le 
{1\over 16}$$
for $L$ sufficiently large.

\\Finally from \equ(6.20) and \equ(6.15)
$$\e^{-(11/4-\eta)}|F_k^{(R)}({\bf u})|\le L^{-1/4}\le {1\over 16}$$
This proves \equ(6.26).
To prove \equ(6.27), take ${\bf u},{\bf u}'\in {\bf B}(1/4)$.
We can then use the Lipshitz estimates of lemma 6.1. 
Note that the initial coupling $g_0$ is held fixed. Then we have
$$\e^{-3/2}|F_k^{(g)}({\bf u})-F_k^{(g)}({\bf u}')|\le 
\sum_{j=0}^{k-1}(2-L^\e)^{k-1-j}\e^{-3/2}
|\tilde \xi(u_j)-\tilde\xi(u_j')|\le$$
$$\le O(1) \e^{5/4-\eta}\Vert {\bf u}-{\bf u}'\Vert
\sum_{j=0}^{k-1}(2-L^\e)^{k-1-j}\le
O(1) \e^{1/4-\eta}\Vert {\bf u}-{\bf u}'\Vert\le
{1\over 2}\Vert {\bf u}-{\bf u}'\Vert$$
Similarly,
$$\e^{-(2-\d)}|F_k^{(\m)}({\bf u})-F_k^{(\m)}({\bf u}')|\le 
L^{-{3+\e\over 2}}\sum_{j=k}^{\io}
L^{-{3+\e\over 2}(j-k)}\e^{-(2-\d)}|\tilde \r(u_j)-\tilde\r(u_j')|\le$$
$$\le L^{-{3+\e\over 2}}O(\log L)\e^{1/2}\Vert {\bf u}-{\bf u}'\Vert\le
{1\over 2}\Vert {\bf u}-{\bf u}'\Vert$$
Finally
$$\e^{-(11/4-\eta)}|||F_k^{(R)}({\bf u})-F_k^{(R)}({\bf u}')|||=
\e^{-(11/4-\eta)}|||U(u_{k-1})-U(u'_{k-1})|||\le
O(1)L^{-1/4}\Vert {\bf u}-{\bf u}'\Vert\le
{1\over 2}\Vert {\bf u}-{\bf u}'\Vert$$
Thus \equ(6.27) has also been proved. This proves lemma 6.3, 
and the proof of theorem 6.2 has now been completed.

\vskip0.5truecm
\\{\it 6.2.  Stable manifold and convergence to fixed point. }

\\Write $E=E_1\times E_2$ with $u\in E$ represented as
$u=(u_1,u_2)$.

\\Here $u_1=(\tilde g, R,{\bf w})$ and $u_2=\m$. $E_1$ and $E_2$
thus represent the contracting and expanding directions for the RG map $f$.

\\Let $p_i,\ i=1,2$, denote the projector onto $E_i$ and $f_i=p_i\circ f$.

\\Note that the norm $\Vert\cdot\Vert$ on $E$ being a box norm
we also have

$$\Vert u\Vert=\sup(\Vert u_1\Vert,\Vert u_2\Vert)$$

\\The following Lemma 6.4, the definition 6.5 of the stable manifold
$W^s$ and our final theorem 6.6 are Irwin's proof of the stable
manifold theorem as presented in appendix 2, chapter 5 of [S]. Part of
Irwin's proof is replaced by theorem 6.2.

\vglue.3truecm
\\{\it Lemma 6.4}

\\Let $u,u'\in B(1/4)$. Then we have
$$\Vert f_1(u)-f_1(u')\Vert\le (1-\e)\Vert u-u'\Vert\Eq(6.30)$$
and, if $\Vert u_2-u_2'\Vert\ge \Vert u_1-u_1'\Vert$ then
$$\Vert f_2(u)-f_2(u')\Vert\ge (1+\e)\Vert u-u'\Vert\Eq(6.31)$$

\vglue.3truecm
\\{\it Proof of lemma 6.4}

\\Because  $u,u'\in B(1/4)$ we can use throughout lemma 6.1.
As always $L$ is sufficiently large and
then $\e$ sufficiently small. First we prove \equ(6.30).
$f_1$ has components $(f_g,f_R,f_{\bf w})$. From \equ(6.4)
$$f_g(u)=(2-L^\e)\tilde g +\tilde\xi(u)$$
Thus using lemma 6.1
$$\eqalign{\e^{-3/2}|f_g(u)-f_g(u')|&\le(2-L^\e)\Vert u-u'\Vert+
\e^{-3/2}|\tilde\xi(u)-\tilde\xi(u')|\cr 
&\le (2-L^\e+O(1)\e^{5/4-\eta})\Vert u-u'\Vert\cr
&\le(1-\e)\Vert u-u'\Vert \cr} $$
for $\e$ sufficiently small.

Since $f_R(u)=U(u)$, we have from lemma 6.1
$$\e^{-(11/4-\eta)}|||f_R(u)-f_R(u')|||\le(1-\e)\Vert u-u'\Vert$$
for $L$ sufficiently large. Finally from the same lemma
$$c^{-1}\Vert f_{\bf w}(u)- f_{\bf w}(u')\Vert\le (1-\e)
\Vert u-u'\Vert$$
These three inequalities prove \equ(6.30).

\\Next we turn to \equ(6.31). In this case by assumption
$\Vert u_2-u_2'\Vert\ge \Vert u_1-u_1'\Vert$ and hence,
since our norms are box norms, we have
$$\Vert u-u'\Vert=\Vert u_2-u_2'\Vert=\e^{-(2-\d)}|\m-\m'|$$
>From \equ(6.5)
$$f_\m(u)=L^{3+\e\over 2}\m+\tilde\r(u)$$
Then, using lemma 6.1,
$$\eqalign{\e^{-(2-\d)}|f_\m(u)-f_\m(u')|\ge
& L^{3+\e\over 2}\Vert u-u'\Vert-
\e^{-(2-\d)}|\tilde\r(u)-\tilde\r(u')| \cr 
& \ge(L^{3+\e\over 2}-O(\log L)\e^{5/2})\Vert u-u'\Vert \cr
& \ge (1+\e)\Vert u-u'\Vert}$$
which proves \equ(6.31) and thus completes the proof of lemma 6.4.
\vglue.3truecm
\\Let $f^k$ be the $k$-fold composition of the RG map $f$.
\vglue.3truecm
\\{\it Definition 6.5}

\\The stable manifold of $f$ is defined by
$$W^s(f)=\{u\in B(1/32):f^k(u)\in B(1/4)\ \forall k\ge 0\}\Eq(6.32)$$
Write the initial points $u$ as $u=(u_1,u_2)$
with $u_1=(\tilde g_0, 0,0)$ and $u_2=\m_0$.
Observe that theorem 6.2 says that there exists for $u\in B(1/32)$
a $u_2$ such that $f^k(u)\in B(1/4)\ \forall k\ge 0$.
We now have
\vglue.3truecm
\\{\it Theorem 6.6}

\\$W^s(f)$ is the graph $\{(u_1,h(u_1)\}$ of a function
$u_2=h(u_1)$ with $h$ Lipshitz continuous with Lip$h\le1$.
Moreover $f|W^s(f)$ contracts distances and hence has a unique 
fixed point which attracts all points of $W^s(f)$.
\vglue.3truecm
\\{\it Proof}

\\To prove the first statement it is enough to prove that if in
$W^s(f)$ we take two points $u=(u_1,u_2)$ and $u'=(u_1',u_2')$
then
$$\Vert u_2-u_2'\Vert\le\Vert u_1-u_1'\Vert\Eq(6.33)$$
because then for a given $u_1$ we would have at most one
$u_2$, and by theorem 6.2 there exists such a $u_2$. This means
that $W^s(f)$ is the graph of a function $h$, $u_2=h(u_1)$,
and moreover 
$$\Vert h(u_1)-h(u_1')\Vert\le\Vert u_1-u_1'\Vert$$
Suppose \equ(6.33) is not true. Then
$$\Vert u_2-u_2'\Vert>\Vert u_1-u_1'\Vert\Eq(6.33.1)$$
Then by \equ(6.31) followed by \equ(6.30) gives
$$\Vert f_2(u)-f_2(u')\Vert\ge (1+\e)\Vert u-u'\Vert>
(1-\e)\Vert u-u'\Vert\ge\Vert f_1(u)-f_1(u')\Vert$$
and hence
$$\Vert f(u)-f(u')\Vert\ge (1+\e)\Vert u-u'\Vert$$
Now
$$\Vert f^2(u)-f^2(u')\Vert=\Vert f(f(u))-f(f(u'))\Vert$$
and by the above and the second part of Lemma 6.4
$$\Vert f^2(u)-f^2(u')\Vert\ge(1+\e)\Vert f(u)-f(u')\Vert
\ge (1+\e)^2\Vert u-u'\Vert$$
By induction we can prove for all $k\ge 0$
$$\Vert f^k(u)-f^k(u')\Vert\ge(1+\e)^k\Vert u-u'\Vert$$
Since $u,u'\in W^s(f)$ the l.h.s. is bounded above by
${1\over 2}$ and hence for $k\rightarrow\io$ we have a
contradiction because $u\ne u'$ under \equ(6.33.1).

\\Hence \equ(6.33) is true and the first statement of theorem
6.6 has been proved.

\\Now we prove that $f|W^s(f)$ is a contraction.
Note that if $u,u'\in W^s(f)$, then
$$\Vert f_2(u)-f_2(u')\Vert\le\Vert f_1(u)-f_1(u')\Vert\Eq(6.34)$$
We can prove this just as we proved \equ(6.33).
Namely assume the contrary and then show in the same way
$$\Vert f^k(u)-f^k(u')\Vert\ge(1+\e)^{k-1}\Vert f(u)-f(u')\Vert$$
The l.h.s. is bounded by ${1\over 2}$ and so as
$k\rightarrow\io$ we get a contradiction because 
$f(u)\ne f(u')$ under the negation of \equ(6.34).
This proves \equ(6.34) which now implies
$$\Vert f(u)-f(u')\Vert=\Vert f_1(u)-f_1(u')\Vert\le (1-\e)
\Vert u-u'\Vert$$
by lemma 6.4, and we are done. Theorem 6.6 has been proved.

\\Note that theorem 6.6 tells us that $\m_0=h(\tilde g_0)=\m_{c}(g_0)$
which defines $\m_c$. If $\tilde g_*=g_*-\bar g$ is
one of the coordinates of the fixed point $u_*$ then
$g_*\ne 0$ since $u_*\in B(1/4)$. In fact the latter implies 

$$|g_*-\bar g|\le {1\over 4}\e^{3/2}$$
and we know $\bar g=O(\e)$ and this excludes $g_*=0$.
So our fixed point is non trivial (non-Gaussian).

\vglue.5truecm

\\{\bf References}



\vglue.3truecm
\\[BDH-est] D. Brydges, J. Dimock and T.R. Hurd: 
Estimates on Renormalization Group Transformation,
Canad. J. Math. 
{\bf 50},
756--793 (1998),  no. 4.

\vglue.3truecm
\\[BDH-eps] D. Brydges, J. Dimock and T.R. Hurd: 
A Non-Gaussian Fixed Point for $\phi^4$ in 4-$\epsilon$ Dimensions,
Comm. Math. Phys.
{\bf 198}, 
111--156 (1998).

\vglue.3truecm
\\[BG] Giuseppe Benfatto and Giovanni Gallavotti:
Renormalization Group,
Princeton University Press, Princeton, NJ,
1995.

\vglue.3truecm 
\\[BY] D. Brydges and H.T. Yau: 
Grad $\phi$ Perturbations of Massless Gaussian Fields,
Comm. Math. Phys.
{\bf 129},
351--392 (1990).

\vglue.3truecm
\\[DH1] J. Dimock and T. R. Hurd:
A Renormalisation Group Analysis of Correlation
Functions for the Dipole Gas,
J. Stat. Phys.
{\bf 66},
1277--1318 (1992).

\vglue.3truecm
\\[DH2] J. Dimock and T. R. Hurd:
Sine-Gordon Revisited,
Ann. Henri Poincar\'e,
{\bf 1}, No 3,
499--541 (2000).

\vglue.3truecm
\\[F] J\"urg Fr\"ohlich:
Scaling and Self-Similarity in Physics,
Birkh\"auser Boston Inc, MA,
1983.

\vglue.3truecm
\\[Gu] G. Guadagni:
University of Virginia Thesis,
in preparation.

\vglue.3truecm
\\[KG] A. N. Kolmogoroff and B. V. Gnedenko,
Limit Distributions of sums of independant random variables,
Addison Wesley, Cambridge,Mass
1954.

\vglue.3truecm
\\[KW] K. G. Wilson and J. Kogut,
The Renormalization Group and the $\epsilon$ Expansion,
Phys. Rep. (Sect C of Phys Lett.),
{\bf 12},
75--200 (1974).

\vglue.3truecm
\\[MS] P. K. Mitter and B. Scoppola:
Renormalization Group Approach to Interacting Polymerised Manifolds,
Comm. Math. Phys.
{\bf 209}
207--261 (2000).

\vglue.3truecm
\\[S] M. Shub:
Global Stability of Dynamical Systems,
Springer-Verlag, New York,
1987.

\ciao